\documentclass[aps,prx,reprint,superscriptaddress]{revtex4-2}

\usepackage{braket}
\usepackage{graphicx}
\usepackage{amsmath}
\usepackage{url}
\usepackage[
    colorlinks = true,
    urlcolor   = blue,     
    linkcolor  = red,      
    citecolor  = green     
]{hyperref}

\begin{document}


\title{Robust and Efficient Quantum Reservoir Computing with Discrete Time Crystal}

\author{Da Zhang}
\author{Xin Li}
    \affiliation{Center for Quantum Technology Research and Key Laboratory of Advanced Optoelectronic Quantum Architecture and Measurements (MOE), \\ School of Physics, Beijing Institute of Technology, Beijing 100081, China}

\author{Yibin Guo}
 \affiliation{Beijing Academy of Quantum Information Sciences, Beijing 100193, China}
\affiliation{Institute of Physics, Chinese Academy of Sciences, Beijing 100190, China}
\affiliation{University of Chinese Academy of Sciences, Beijing 101408, China}
    
\author{Haifeng Yu}
\affiliation{Beijing Academy of Quantum Information Sciences, Beijing 100193, China}
\affiliation{Beijing Key Laboratory of Fault-Tolerant Quantum Computing}

\author{Yirong Jin}
\affiliation{Beijing Academy of Quantum Information Sciences, Beijing 100193, China}
\affiliation{Beijing Key Laboratory of Fault-Tolerant Quantum Computing}
\affiliation{Xoherence Co., Ltd., Beijing 100193, China}
  
\author{Zhang-Qi Yin}
    \email{zqyin@bit.edu.cn}
    \affiliation{Center for Quantum Technology Research and Key Laboratory of Advanced Optoelectronic Quantum Architecture and Measurements (MOE), \\ School of Physics, Beijing Institute of Technology, Beijing 100081, China}

\date{\today}

\begin{abstract}
The rapid development of machine learning and quantum computing has placed quantum machine learning at the forefront of research. However, existing quantum machine learning algorithms based on quantum variational algorithms face challenges in trainability and noise robustness. In order to address these challenges, we introduce a gradient-free, noise-robust quantum reservoir computing algorithm that harnesses discrete time crystal dynamics as a reservoir. 
We first calibrate the memory, nonlinear, and information scrambling capacities of the quantum reservoir, revealing their correlation with dynamical phases and non-equilibrium phase transitions. We then apply the algorithm to the binary classification task and establish a comparative quantum kernel advantage. For ten-class classification, both noisy simulations and experimental results on superconducting quantum processors match ideal simulations, demonstrating the enhanced accuracy with increasing system size and confirming the topological noise robustness.  Our work presents the first experimental demonstration of quantum reservoir computing for image classification based on digital quantum simulation. It establishes the correlation between quantum many-body non-equilibrium phase transitions and quantum machine learning performance, providing new design principles for quantum reservoir computing and broader quantum machine learning algorithms in the NISQ era.

\end{abstract}

\maketitle

\section{Introduction}

Machine learning has achieved transformative breakthroughs in the past decade \cite{lecun2015learn,wang2023learn}, delivering superhuman performance in diverse domains, from image classification \cite{krizhevsky2017alexnet,he2016resnet} and prediction of protein structures \cite{senior2020protein} to natural language processing \cite{vaswani2017attention}. Parallel to machine learning, rapid progress in quantum hardware has brought noisy intermediate-scale quantum (NISQ) computation \cite{preskill2018nisq} from theoretical concepts to experimental reality. Leading platforms, including superconducting circuits \cite{acharya2024google,kim2023evidence}, trapped ions \cite{guo2024500} and neutral atoms \cite{bluvstein2024logical}, scale to hundreds of physical qubits while demonstrating improved gate fidelity beyond the fault-tolerant threshold. This convergence of capabilities establishes quantum machine learning (QML) as a promising frontier for advancing next-generation computing paradigms  

Quantum machine learning algorithms have demonstrated computational enhancements in multiple aspects, including quantum simulation for chemistry and materials science \cite{kandala2017vqe}, quantum and classical classification \cite{cong2019quantum,herrmann2022qcnn,hur2022qcnnc} and combinatorial optimization \cite{farhi2014qaoa}. However, most of these implementations mainly rely on variational quantum algorithms (VQAs) trained using classical optimization techniques. These algorithms face three major difficulties when scaling up. First, unlike classical deep learning, computing gradients for quantum systems is resource-intensive \cite{pan2023deepgradient}, both theoretically and experimentally. Furthermore, the existence of barren plateaus \cite{mcclean2018bpqnn,anschuetz2022entanbps} and local minima \cite{bittel2021vqanp} leads to gradients that become exponentially suppressed with increasing quantum processor scale, significantly increasing the resources required for convergence to a global minimum. Finally, noise typically scales with the size of the system and the depth of evolution, causing the converged results to deviate from the ideal results.

To overcome these limitations, quantum reservoir computing \cite{qrc17kohei,qrc19kohei,ghosh2019qrcp,mujal2023qrcweak,ghosh2021qrccircuits,mujal2021qrcreview,ghosh2019qrcp}  (QRC) has gained attention as an alternative algorithm. QRC is an algorithm developed within the framework of the classical reservoir computing (CRC) theory \cite{jaeger2001crc,maass2002crc,tanaka2019crc,yan2024crc}. In CRC, the hidden layer of a traditional neural network is replaced by a dynamic reservoir consisting of a stochastic sparse network. The behavior of the reservoir is controlled by adjusting the global parameters, including the size of the network, the spectral radius, and the leak rate. This mechanism effectively achieves computational acceleration of the training process. Since the reservoir is fixed in training once it is generated, the dynamics of the reservoir must satisfy three basic requirements: possessing nonlinear response capacity to fully unfold the feature space; exhibiting the fading memory property for dynamic processing of time series; and forming a high-dimensional space projection to guarantee linear separability of input features. Quantum many-body systems exhibit unique physical advantages in satisfying these properties. Such systems naturally provide an exponentially scaled dimension through many-body entanglement, nonlinear response induced by dynamical evolution, and the fading memory property arising from eigenstate thermalization \cite{deutsch2018eth,martinez2021qrcdynamical}. By leveraging the dynamics of quantum systems as computational resources, QRC inherently avoids gradient-based optimization in the quantum aspect and offers enhanced scalability. 

Recent advances in QRC have witnessed the realization of diverse physical models. QRC based on many-body systems naturally forms recurrent neural network architectures through inter-qubit interactions, making them suitable for time series processing \cite{xia2023qrcconfigured,bravo2022quantumtimex}. Additional implementations include reconstructing state properties via quantum random walks \cite{suprano2024expproper}, and performing image classification using various mechanisms, including Ising-like interactions \cite{de2025qelmimage}, neutral-atom interactions \cite{kornjavca2024analog}, and Gaussian boson sampling \cite{cimini2025gbs,gong2025gbs}. Although these approaches successfully reduce quantum parameter optimization requirements, most work, especially those involving quantum many-body dynamics, remains to be theoretically validated without comprehensive analysis on noise effects in practical systems. These advances raise the critical question: "What fundamental characteristics define an optimal quantum reservoir?" For time series processing, the memory, nonlinearity and information scrambling capacities of the quantum reservoir were comprehensively calculated, and it was found that the system has the strongest information processing capabilities when it is near the boundary of the dynamical phase transition \cite{martinez2021qrcdynamical,xia2022qrcdy}. We explore this question further from the perspective of realizing image classification with quantum many-body models.

In this paper, we propose a gradient-free, noise-robust QRC algorithm employing the many-body localized discrete time crystal \cite{dtcrigid,dtcrmp} (MBL-DTC, hereafter referred to as DTC) evolution as a quantum reservoir. The unique non-equilibrium phase transitions and long-lived temporal order in DTC provide a theoretically grounded framework for reservoir design, enabling robust and efficient information processing. We first calibrate the memory, nonlinear, and information scrambling capabilities of a DTC model using time series processing tasks. The results of these tasks establish a direct correlation between the dynamical phases and the information processing power of QRC. Then, by maximizing the performance difference between quantum and classical algorithms through kernel methods, we reveal potential comparative quantum kernel advantages in the binary classification task. 
Through the numerical simulation of the ten-class classification task, we show that the image classification accuracy also exhibits nonlinear variation characteristics in different phases of matter regions.

Before testing the algorithm in experiments, we numerically simulate the effects of both the depolarizing noise and the hardware-related thermal relaxation noise. The noise robustness and scalability of the proposed algorithm have been tested and verified numerically. Based on the above results, the experiments conducted on the Quafu quantum cloud platform further confirmed the noise robustness, the relationship between the non-equilibrium phase transition and the image classification accuracy. It is also observed that the classification accuracy increases with increasing system scale.

\section{Quantum Reservoir Computing Framework}

\begin{figure*}
\includegraphics[width=\linewidth]{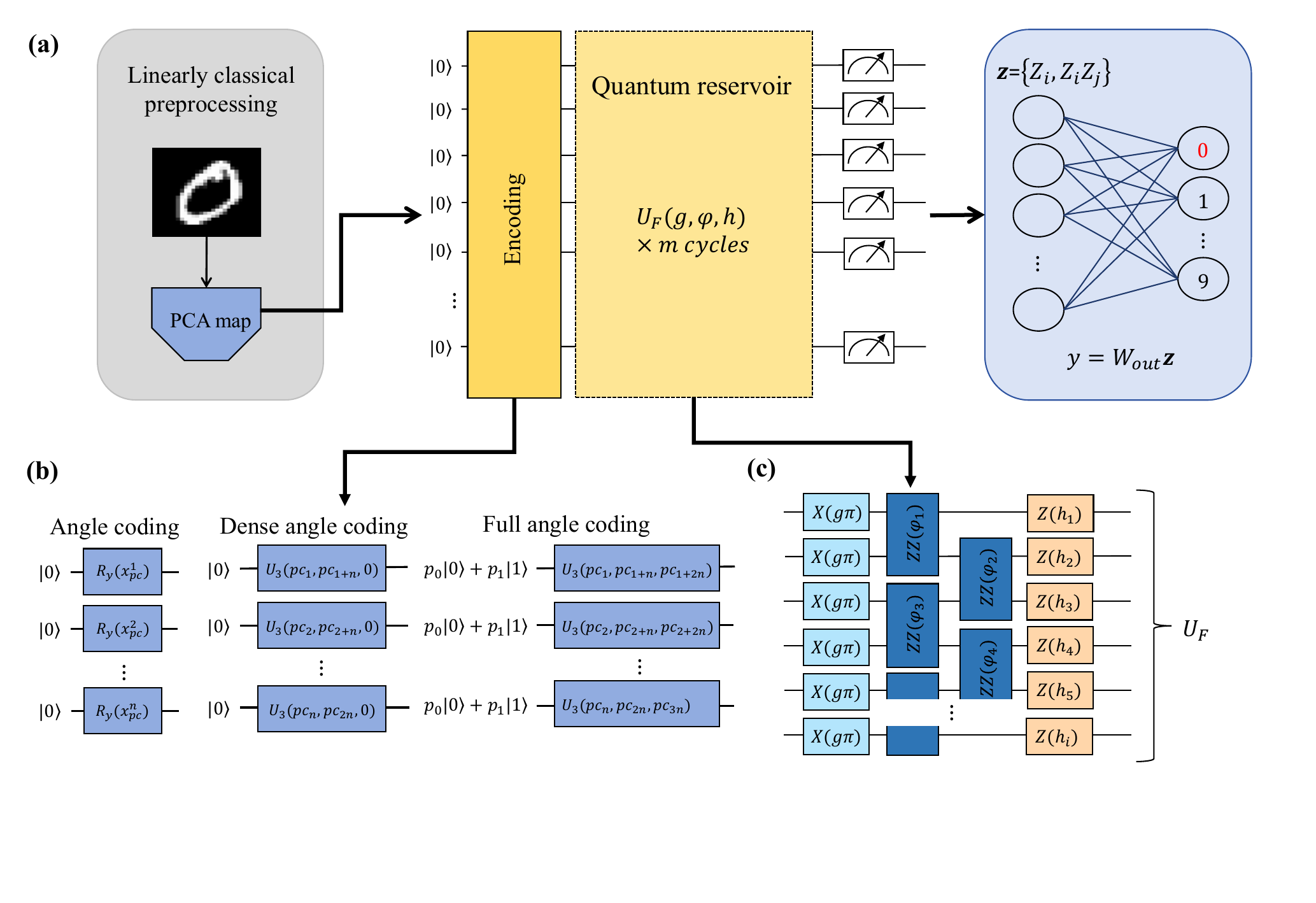}
\caption{Schematic diagram of quantum reservoir computing with discrete time crystal.
    (a)Workflow of quantum reservoir computing, including PCA-based dimensionality reduction of image data,  data encoding, quantum evolution, measurement, and classical training.
    (b) Data encoding schemes: ideal cases employ angle/dense angle encoding, while the experimental implementation uses full-angle encoding due to the presence of state preparation error. 
    (c) Circuit implementation of a single layer of $U_F$, including imperfect flip $X(g\pi)$, $ZZ(\phi_i)$ interactions, and transverse field $Z(h_i)$ terms.
    \label{fig:qrc}}
\end{figure*}
In this section, we briefly introduce the central idea of QRC. As shown in Fig.~\ref{fig:qrc}(a), the protocol includes three main components: (1) classical input processing and encoding, (2) quantum reservoir dynamics, and (3) measurement and training. This section details the classical pre- and post-processing procedures, while the implementation of the quantum reservoir using DTC is presented in detail in Section \ref{dtc}. 

We focus on supervised classification tasks, where each input sample $\mathbf{x} \in X$ has a label $y\in Y$. The training set is $S_{N}=\left \{ (\mathbf{x}_{1},y_{1}),\dots,(\mathbf{x}_{N},y_{N}) \right \} $, where $N$ represents the size of the training set. The goal of classification is to learn a function $f(\mathbf{x})$ that accurately predicts the labels of unseen samples. Unlike most existing QRC algorithms that rely on analog simulation, our quantum reservoir algorithm based on discrete time crystals (DTC-QRC) is based on digital simulation and is therefore applicable to most universal quantum processors.  

\subsection{Input Processing and Encoding Schemes \label{encoding}}

To bridge the dimensionality gap between classical image data and quantum processors, we employ principal component analysis (PCA) for dimensionality reduction. This linear transformation projects high-dimensional data into a lower-dimensional space while preserving essential features with minimal distortion. The resulting principal components $\mathbf{x}^{pc}$ are encoded in the quantum states $\phi(\mathbf{x}^{pc})$ using single-qubit gates. Encoding data in the initial state instead of evolution ensures that the parameters of the Hamiltonian are within a specific interval, thus giving a convenient way to explore the correlation between the properties and the computational capacity of the quantum reservoir.

As shown in Fig.~\ref{fig:qrc}(b), we use three encoding schemes: two for simulation and one for experiment. The first is angle encoding, which is the standard method for time series processing. Angle coding can be achieved by applying a $RY(\theta)$ gate to $\ket{0}$
\begin{equation}
Ry(x^{pc}_{i})\ket{0}=\cos(\frac{x^{pc}_{i}}{2})\ket{0}+\sin(\frac{x^{pc}_{i}}{2})\ket{1}.
\end{equation}

In order to encode more classical data per qubit, dense angle encoding is used \cite{larose2020robustencoding}. In the ideal case, the quantum state is prepared on the Bloch sphere and is independent of $x_{2n+i}^{pc}$
\begin{equation}
    U_{3}(x_{i}^{pc},x_{n+i}^{pc},x_{2n+i}^{pc})\ket{0}=\cos(\frac{x_{i}^{pc}}{2})\ket{0}+e^{ix_{n+i}^{pc}}\sin(\frac{x_{i}^{pc}}{2})\ket{1}.
\end{equation}

However, in real experiments, there are errors in the preparation of quantum state. A state that should be prepared as $\ket{0}$ has the probability $p_{1}$ of being prepared as $\ket{1}$
\begin{eqnarray}
&U_3(\mathbf{x}^{pc}) (p_0\ket{0} + p_1\ket{1}) 
= p_0 \bigl[ \cos\tfrac{x^{pc}_{i}}{2}\ket{0} + e^{i x^{pc}_{n+i}}\sin\tfrac{x^{pc}_{i}}{2}\ket{1} \bigr] \nonumber \\
&+ p_1 \bigl[ {-}e^{i x^{pc}_{2n+i}}\sin\tfrac{x^{pc}_{i}}{2}\ket{0} + e^{i(x^{pc}_{n+1}+x^{pc}_{2n+i})}\cos\tfrac{x^{pc}_{i}}{2}\ket{1} \bigr].
\end{eqnarray}

Our experiments show that full $U_{3}$ encoding significantly improves classification accuracy under noise.

Given the input principle component $\left \{ \mathbf{x}^{pc} \right \} $, the angle encoding evolution can be expressed as 

\begin{equation}
\ket{\phi(\mathbf{x}^{pc})}=\bigotimes_{i=1}^{n}(Ry(x_{i}^{pc})\ket{0}),
\end{equation}

and the dense angle and full angle coding

\begin{equation}
\ket{\phi(\mathbf{x}^{pc})}=\bigotimes_{i=1}^{n}(U_{3}(x^{pc}_{i},x^{pc}_{n+i},x^{pc}_{2n+i})(p_{0}\ket{0}+p_{1}\ket{1})).
\end{equation}

\subsection{Measurement and Classification}
After encoding, the system evolves under a time-periodic unitary evolution $U_{F}$ with period T, thereby simulating the dynamics of a discrete time crystal. This quantum reservoir induces nonlinear data transformations and exhibits fading memory effects, enabling classically intractable dynamics as the system scales up. 

To harness the temporal data processing capabilities of the quantum reservoir, the evolution cycles $m$ are divided into four equal segments. Following each segment of $\frac{m}{4}$ cycles, projective measurements are performed, which yield the expectations of local operators along the z-axis $\mathbf{z}^{(k)}=\left \{ Z_{i}^{k},Z_{i}^{k}Z_{j}^{k} \right \}$ as features. Specifically, these are given by
\begin{eqnarray}
        Z_{i}^{k} &= \bra{\phi(x^{\mathrm{pc}})}(U_{F}^{\dagger})^{k} \sigma_{z}^{i}(U_{F})^{k}\ket{\phi(x^{\mathrm{pc}})} \\
    Z_{i}^{k}Z_{j}^{k} &= \bra{\phi(x^{\mathrm{pc}})}(U_{F}^{\dagger})^{k} \sigma_{z}^{i}\sigma_{z}^{j}(U_{F})^{k}\ket{\phi(x^{\mathrm{pc}})}.
\end{eqnarray}
The full output quantum features are then constructed as
\begin{equation}
    \mathbf{Z} =\left \{  (\mathbf{z}^{(k)})^{T}\right \}.
\end{equation}

Classification is performed by feeding the features directly into a classical single-layer fully connected network (SLN) without using the active function. The layer implements the linear transformation is

\begin{equation}
\mathbf{y} = W_{out}\mathbf{z} + \mathbf{b}.
\end{equation}
We stress that although this simple classical method helps us study the natural properties of the quantum reservoir, it does not show the full potential of quantum reservoir computing.  The enhanced performance achievable with multilayer fully connected networks is shown in \ref{sample}.

\section{Discrete Time Crystal as a Quantum Reservoir \label{dtc}}
In this section, we introduce the discrete time crystal based quantum reservoir computing and calibrate the properties of the dynamics. The DTC represents a novel non-equilibrium phase characterized by spontaneous breaking of discrete time translation symmetry under periodic driving, manifesting robust subharmonic dynamical responses (e.g. period doubling) \cite{Khemani2019}. 
This phenomenon occurs for an extended parameter range and robustness to the choice of initial state, making it suitable for implementation on the NISQ platforms \cite{ippoliti2021manydtcnoise}. DTC has been observed on a variety of experimental platforms, ranging from trapped ions \cite{zhang2017trapdtc}, superconducting circuits \cite{frey2022superdtc} to nitrogen-vacancy centers \cite{randall2021nvdtc}. Recent work demonstrates that DTC quantum circuits can mitigate barren plateaus in variational quantum eigensolvers when employed as ansätze \cite{DTC-VQE2025}. This suggests that quantum machine learning implementations that use DTC dynamical evolution merit systematic investigation. 

The dynamical evolution of DTC is governed by a unitary time evolution operator $U_{F}$ \cite{mi2022time}.
\begin{equation}
U_F = \begin{cases}
U_{\text{flip}} = e^{-\frac{i}{2}\pi g\sum_{i}X_{i}}, & 0 \le t \le t_{1} \\
U_{\text{int}} = e^{-\frac{i}{4}\varphi_{i} Z_{i}Z_{i+1}}, & t_{1} \le t \le t_{2} \\
U_{\text{onsite}} = e^{-\frac{i}{2}\sum_{i}h_{i}Z_{i}}, & t_{2} \le t \le T 
\label{dtceq}
\end{cases}
\end{equation}
where angles $\varphi_{i}$ are sampled randomly from $[-1.5\pi,-0.5\pi]$, and disorder strength $h_{i}$ are sampled randomly from $[-\pi, \pi]$. The circuits implementing $U_{F}$ are shown in Fig.~\ref{fig:qrc} (c), and achieve an interacting Ising model that is periodically flipped by $\pi g$ about the $x$ axis. As $g$ is tuned from $0$ to $1$, the system undergoes dynamical phase transitions from the many body localization non-time-crystal (MBL-NTC, hereafter referred to as NTC) phase, to the thermal phase, and finally to the DTC phase. 

Many-body localization is a dynamical phase of matter that occurs in strongly disordered, interacting quantum many-body systems. Prevents the system from reaching thermal equilibrium by suppressing long-range transport, preventing entanglement spreading, and preserving memory of initial conditions. Achieving a DTC phase requires many-body localization \cite{pal2010mbl} to prevent the drive from heating the system to an infinite temperature. However, in the thermal phase, even in the absence of decoherence effects, the local feature of the system rapidly loses its memory of the initial state, and the expectation of the local observables evolves to the thermal equilibrium value. Consequently, information about the initial state cannot be recovered through local measurements. 

\subsection{Absolute Stability in Discrete Time Crystal \label{stability}}
In DTC phase, eigenstates exhibit a characteristic Schrödinger cat pairing structure with energy gaps of $\pi$. This $\pi$ pairing is robust to weak Ising symmetric perturbations and all $T$ periodic weak perturbations \cite{Absolutestability,spinglass2016,Khemani2019}. Compared to other non-equilibrium phases, the DTC stability is unique and constitutes a broad, topologically protected phase. This topological protection makes the system intrinsically resistant to a wide range of non-ideal and control errors (coherent errors) in experiments. Recent work has further revealed the existence of nonlocal Majorana edge modes for $\pi$ pairing, and that any multi-qubit Pauli operator with overlap with MEMs is resistant to integrability
and symmetry-breaking fields as well as dephasing noise \cite{edgerobust,dtcrobust}. The quantum reservoir in the DTC phase also exhibits enhanced stability against decoherence. Another manifestation of stability is for DTC phase, even when quenched from generic short-range correlated initial states, long-time evolution still exhibits characteristic subharmonic oscillations \cite{Absolutestability}. As discussed in \ref{encoding}, the initial state is prepared using single-qubit gates, which only produce states with short-range correlations. As a result, DTC imposes less stringent requirements on initial state preparation.

\subsection{Correlation of Dynamical Phase Transitions with Quantum Reservoir Capacities}

Only linear pre- and post-processing are implemented for image classification, which necessitates that the quantum reservoir itself processes short-term memory and sufficient nonlinearity. Moreover, for quantum many-body systems, the initially encoded information needs to diffuse through the system at an appropriate rate. In the following, we demonstrate that quantum systems with dynamical phase transition boundaries can effectively balance the trade-offs among these essential requirements, highlighting their potential as efficient quantum reservoirs.

\begin{figure}
    \centering
    \includegraphics[width=0.98\linewidth]{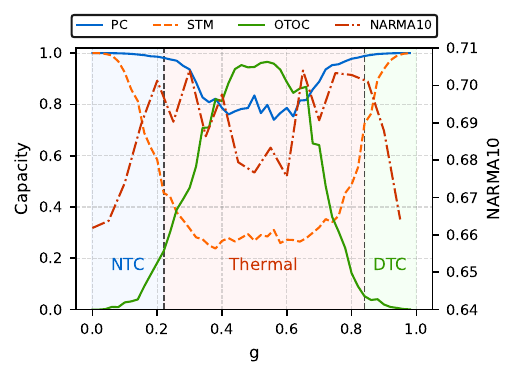}
    \caption{Capacities of discrete time crystal based quantum reservoir computing on the short-term memory, parity check, NARMA10 tasks and information scrambling.  }
    \label{fig:cap}
\end{figure}

\subsubsection{Short-Term Memory and Nonlinear Capacity}
We calibrate short-term memory and nonlinear capabilities using established temporal processing tasks \cite{xia2022qrcdy,martinez2021qrcdynamical}. In our DTC-QRC, the input is injected into the same qubit every time step $t$ $\rho_{s_{k}}=\ket{\psi_{s_{k}}}\bra{\psi_{s_{k}}}$, where $\ket{\psi_{s_{k}}}=\sqrt{1-s_{k}}\ket{0}+\sqrt{s_{k}}\ket{1}$ and $s_{k}$ are the input time series data.  At time t, the density matrix is interrupted by the input as $\rho(t)=\rho_{s_{k}}^{i}\otimes Tr_{i}[U_{F}\rho(t-1)U_{F}^{\dagger}]$. Measurements are performed at the end of each time step given $\mathbf{z}$. The measurement results are then used to train a linear regression model $y(t)=\sum W_{out}\mathbf{z}+B$. The weights $W_{out}$ and the bias $B$ are optimized to minimize $\left \| y-y^{\ast} \right \|$, where $y^{\ast}$ is the target time series. The memory capacity is calibrated through the short memory task \cite{qrc17kohei}, where the target time series is $y^{\ast}_{k}=s_{k-\delta}$ with $\delta$  representing the time delay. Nonlinear capacity is calibrated by the parity check task, which targets series defined as $y^{\ast}_{k}=(\sum_{m=0}^{\delta}s_{k-m}) mod 2$. In both tasks, the input series is binary $s_{k} \in \left \{ 0,1 \right \} $.

The nonlinear auto-regressive moving average (NARMA) model is widely used for recurrent neural network benchmarks. We test DTC-QRC's information processing capacity using NARMA10 where
\begin{equation}
    y_{k}=0.3y_{k-1}+0.05y_{k-1}(\sum_{j=1}^{10}y_{k-j})+1.5s_{k-10}s_{k-1}+0.1
\end{equation}
For this task, the quantum reservoir needs to learn $y_{k}$ from the binary random input $s_{k}$. 

For all three tasks,the capacity is quantified by a normalized covariance,
\begin{equation}
    C=\frac{\mathrm{cov}(y,y^{\ast})}{\sigma(y)\sigma(y^{\ast})}.
\end{equation}
The normalized covariance $C$ quantifies how well the reservoir predictions match the target time sequence. Each task consists of $1000$ time steps, with a $9:1$ ratio for the training and testing sets. 

\subsubsection{Information Scrambling Capacity}

The information scrambling capacity is given by one minus the average of the out-of-time-ordered correlator (OTOC) over all qubits, taken for information that is initially injected into the first qubit \cite{larkin1969cotoc,li2017otoc}
\begin{equation}
    C_{in}(t)=1-\frac{1}{n-1}\sum_{i=2}^{N}\left \langle X_{1}^{\dagger}(t)X_{i}^{\dagger}X_{1}(t)X_{i}  \right \rangle,
\end{equation}
with $X_{1}(t)=(U_{F}^{k})^{\dagger}X_{1}U_{F}^{k}$ represents the time-evolved Pauli operator under the Floquet unitary. The OTOC $\left \langle X_{1}^{\dagger}(t)X_{i}^{\dagger}X_{1}(t)X_{i}  \right \rangle $ captures the information spreading from the first qubit to the rest. 

As shown in Fig.~\ref{fig:cap},  increasing the imperfect flip strength $g$ shows distinct dynamical regimes characterized by nonmonotonic evolution of reservoir computing capabilities.  Both short-term memory and nonlinear processing capacities exhibit a characteristic decrease followed by recovery, while the quantum information scrambling capacity displays the opposite trend. $ C_{in}(t)$ remains suppressed in both the NTC and DTC phases, reflecting the preservation of quantum information and restricted entanglement growth.  For the NARMA10 benchmark task, optimal performance emerges when these three computational capacities achieve balance at the phase transition boundaries. 

\section{Testing Comparative Quantum Kernel Advantage}

\begin{figure}
    \centering
    \includegraphics[width=0.98\linewidth]{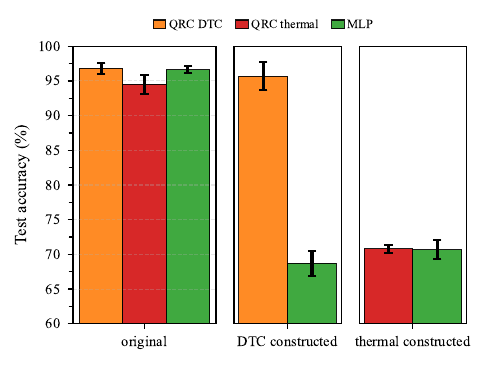}
    \caption{Comparative quantum kernel advantage with QRC. The left panel shows the test classification accuracy of simulated QRC in the DTC and thermal phases, as well as an MLP, on the original 3/8 binary MNIST classification task. 
The two panels on the right depict the kernel geometry-based contrastive tasks for the DTC and thermal phases, respectively. }
    \label{fig:kernel}
\end{figure}

We first show the effectiveness of the QRC algorithm using the MNIST $3/8$ binary classification task. As shown in Fig.~\ref{fig:kernel},  simulation results at the DTC phase transition critical boundary (\textit{g}=0.84) and in the thermal phase (\textit{g}=0.5) are compared to a classical multilayer perceptron (MLP) with four fully connected layers and dropout.  The QRC with $12$ qubits at criticality achieves an accuracy comparable to the MLP, while its performance degrades slightly in the thermal phase. 

To further explore potential quantum enhancements of the QRC algorithm, we employ the recently proposed geometric kernel difference approach \cite{huang2021power,wang2021towards}, which treats the QRC as a kernel mapping that transforms classical data into quantum feature space. For quantum ($K_{Q}$) and classical ($K_{C}$) kernels, when their geometric behavior differs significantly,  we can transform the dataset labels to amplify the performance difference between kernels.

The geometric quantity $g_{CQ}$ quantifies potential quantum performance difference through kernel comparison and is defined as: 
\begin{equation}
    g_{CQ}=\sqrt{\left \| \sqrt{K_{C}}(K_{Q})^{-1}\sqrt{K_{C}} \right \|_{\infty}},
\end{equation}
where $\left \|  \right \|_{\infty}$ is the spectral norm of the resulting matrix. We can construct a contrastive data set. When $g_{CQ}$ is large , there exists a function where quantum models exhibit superior prediction accuracy.

To rigorously characterize the quantum-classical separation, we employ the radial basis function (RBF) kernel,
\begin{equation}
    K(\mathbf{x}_{i},\mathbf{x}_{j})=\exp (-\gamma \left \| \mathbf{x_{i}}-\mathbf{x}_{j}  \right \|^{2}).
\end{equation}
The new labels of the contrastive datasets are constructed through reassigning the new label to the data according to the values of $y^{\ast}=\sqrt{K_{Q}}\left \| \sqrt{K_{C}}(K_{Q})^{-1}\sqrt{K_{C}} \right \|_{\infty}$. For a binary classification task, the new labels can be constructed by the median binarization of $y^{\ast}$

\begin{equation}
    y^{re}=\left\{\begin{matrix}
1  &  \text{if} \quad y^{\ast}_{i} > \mathrm{median}(y^{*})\\
0  &  \text{if} \quad y^{\ast}_{i} < \mathrm{median}(y^{\ast})
\end{matrix}\right .
\end{equation}

Through optimal hyperparameter selection ($\gamma$), the quantum-classical classification accuracy difference at the dynamical phase transition boundary (g=0.84) increases from 0.2\% to 27\% after constructing the confrontational dataset. In contrast, this difference remains negligible in the thermal phase even after dataset transformation.   This phase-dependent enhancement confirms our geometric kernel distance analysis, demonstrating that the method effectively reveals the computational capabilities of QRC systems. The results provide evidence that QRC can achieve quantum kernel advantages, but only when operating within appropriately dynamical phases that preserve quantum geometric properties.

\section{Multiclass Classification and Noise Robustness}

\begin{figure*}[!t] 
\centering
\includegraphics[width=1.96\columnwidth]{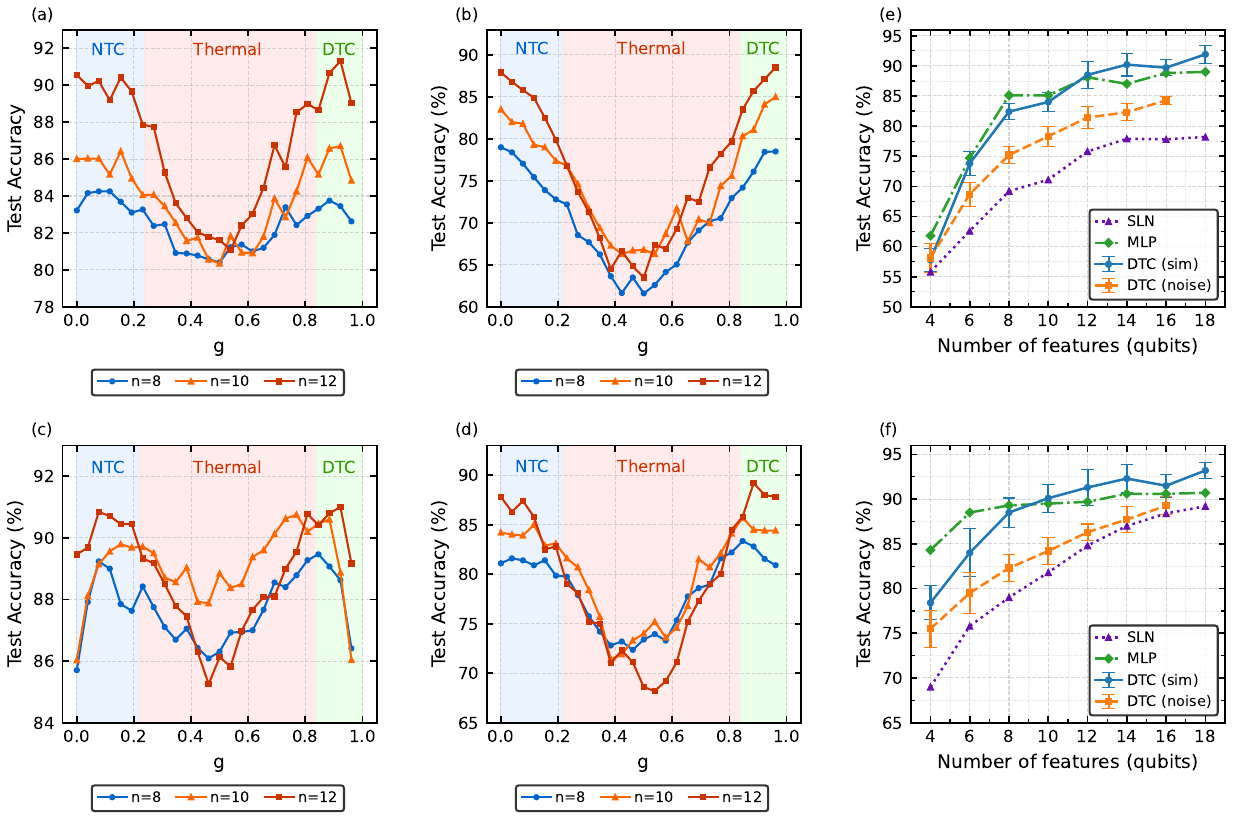}
\caption{
Model performance in different dynamical phases and system scales. (a–d) Phase-dependent performances: (a,b) with angle encoding; (c,d) with dense angle encoding. Ideal results are shown in (a,c), while corresponding noisy results under depolarizing noise (single-qubit $p=0.001$, two-qubit $p=0.02$) appear in (b,d). Scaling behavior with system size is characterized in (e,f) using experimentally relevant thermal relaxation noise. 
\label{fig:simres}}
\end{figure*}

Having established the quantum kernel advantage in binary classification, we extend our analysis to multiclass tasks. Fig.~\ref{fig:simres} shows the QRC classification accuracy on the ten-class MNIST dataset under ideal and noisy simulations. We used 1200 training and 200 test samples to balance computational cost with performance while matching experimental constraints.  The evolution depth is set to m=12. Training epochs for the classical layer are optimized via grid search, and the results are averaged over 10 independent realizations of random parameters $ \left \{  \varphi_{i},h_{i} \right \} $. 

Fig.~\ref{fig:simres} (a) and (b) show the simulation results based on the angle coding. With this encoding scheme, the test accuracy increases as the thermal phase transitions towards either the NTC phase or the DTC phase, and slightly decreases at both terminals. Fig.~\ref{fig:simres} (c) and (d) present results for dense angle encoding, which have a similar trend to angle encoding, but the test accuracy peaks in two phase transition boundaries. The emergence of these phenomena can be explained by an antagonistic mechanism. The quantum reservoir needs to preserve the memory of the initial encoded information, but also need to make it scrambling through the system at a suitable rate to map it to a higher dimensional space and provide sufficient nonlinearities.  According to the capacities calibrated in Fig.~\ref{fig:cap}, both NTC and DTC protect the initial information, but equally inhibits information scrambling and nonlinearities. As a result, accuracy degradation occurs near $g=0$ or $g=1$ (complete protection of initial information in the absence of noise )  for both encoding methods. For dense angle coding, the drop is particularly large near the two terminals as more information is encoded in a single qubit, requiring greater information scrambling capacity. This antagonism is just balanced at boundaries of the non-equilibrium phase transition. The results highlight the critical role of phase transitions in tuning the computational power of the reservoir for classification tasks. 

To demonstrate the scalability of the DTC-QRC,  Fig.~\ref{fig:simres}(e) and (f) show classification accuracy as a function of the quantum reservoir size. The accuracy improves with system size up to 18 qubits, providing preliminary evidence of scalability.  After adding the quantum reservoir, the classification accuracy is significantly higher than that of a SLN and comparable to that of a MLP. The architecture of the networks are detailed in Table \ref{tab:nn}.  This suggests that DTC-QRC can serve as an efficient quantum classifier when utilizing a suitable phase of matter.  

\begin{figure*}
\includegraphics[width=\linewidth]{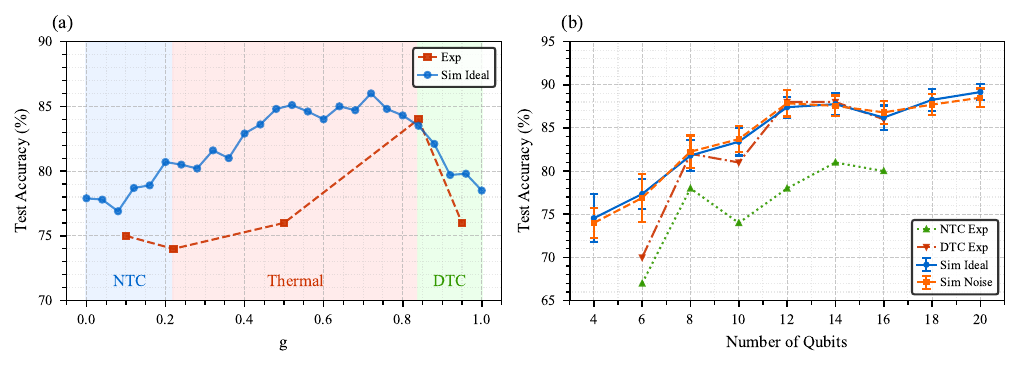}
\caption{Experimental results on cloud quantum platform.
    (a) Relationship between non-equilibrium phase transitions and classification accuracy;
    (b) Qubit scaling performance in DTC and NTC phase transition boundaries. 
    \label{fig:exp}}
\end{figure*}

\subsection{Effect of Noise}
We present a numerical analysis of the DTC-QRC algorithm's noise robustness, focusing on three dominant noise sources: decoherence noise, quantum gate imperfections (coherence noise), and state preparation and measurement (SPAM) noise. 

In both NTC and DTC phases, the suppression of long-range entanglement growth serves as a fundamental mechanism that prevents the onset of global error cascades, ensuring that decoherence effects do not diffuse exponentially with increasing system size \cite{ippoliti2021manydtcnoise}. Due to the extended parameter range and absolute stability in the DTC phase, it is robust to small deviations of rotation angles and unwanted couplings. This "forgiving" parameter space reduces the need for precision alignment and control, making DTC implementations more experimentally accessible. Furthermore, SPAM noise and all other residual noise can be mitigated through the implementation of a supervised learning protocol applied to the final measurement outcomes. This joint mechanism significantly mitigates the effect of noise on model performance.

In our investigation of the relationship between quantum phases and classification accuracy, we employed a depolarizing noise model. As shown in Fig.~\ref{fig:simres} (b) and (d), accuracy drops by up to nearly $20\%$ in the thermal phase, but $3\%$ in NTC and DTC phases, demonstrating the noise robustness of the algorithm.  The presence of noise destroys the protection of MBL and facilitates the diffusion of information, driving the emergence of optimal outcomes deeper into both the NTC and DTC phases.  

For scalability tests, we employ a thermal relaxation noise model that closely matches experimental conditions.  The details of the noise model and the basis for the selection of the noise strength are discussed in Appendix \ref{noisemodel}.   As shown in Fig.~\ref{fig:simres} (e) and (f), the algorithm’s performance under noise is slightly worse to the ideal case, yet it continues to improve with increasing system size. This is mainly because at current scales, quantum many-body effects become more pronounced with increasing scale, leading to stronger noise resistance. Moreover, as the number of qubits increases, the measured expectation values of single-body and two-body operators increase polynomially. The corresponding adjustable parameters in SLR also increase, providing enhanced noise resilience.  

\section{Experimental Results}

We experimentally demonstrate our algorithm on the \textit{'Baihua'} superconducting quantum processor provided by \textit{Quafu Superconducting Quantum Computing}~\cite{QuafuSQC}. \textit{'Baihua'} contains $156$ qubits. Experiments are done in a chain with maximum of $16$ qubits. The median single-qubit and two-qubit gate error rates of this processor are $8\times10^{-4}$ and $2\times 10^{-2}$. Due to experimental resource constraints, we set the total Floquet cycles $m=4$, and set the number of train samples and test samples as 500 and 100. The shots in experiments are set as $1024 \times n/2$ where $n$ is the number of qubits. The parameters of the circuits are chosen randomly in each experiment. We also use full angle coding to encode more principal components and improve the noise robustness of the algorithm.

\subsection{Performance Enhancement at the Phase Transition Boundary}
We then applied the 10-class MNIST classification task to experimentally investigate the correlation between classification accuracy and non-equilibrium phase transitions. Specifically, we selected system parameters which correspond to different phases and phase transition boundaries, given by $g =\left \{ 0.1,0.22,0.5,0.84,0.95 \right \} $, using a ten qubit quantum reservoir. The experimental results in Fig.~\ref{fig:exp}(a) reveal that the peak of classification accuracy is achieved near the phase transition boundary between the thermal and DTC phases. In contrast, the phase transition boundary between the thermal and NTC phases does not exhibit a comparable enhancement in classification performance, as confirmed by numerical simulations. 

\subsection{Qubit Scaling Performance}
Based on phase-dependent performance analysis,  we selected $g=0.84$ to further demonstrate the scalability of the algorithm in our experiments. Fig.~\ref{fig:exp}(b) shows that the experimental results are fitted to the simulation results and achieve maximum accuracy $88\%$ using 12 and 14 qubits. 
In experiments, the controlled phase gate is decomposed as $rzz(\theta)(k,k+1)=\mathrm{CNOT}(k,k+1)rz(\theta)(k)\mathrm{CNOT}(k,k+1)$. There are up to $78$ CNOT gates in the experiments. The median error rate of the CNOT gate in experimental equipment is $2\%$,  which is enough to destroy the quantum information. However, our algorithm can still achieve results similar to those of classical algorithms. When the system was scaled up to 16 qubits, the performance began to decrease. And due to the limitations of the experimental system, larger experiments were not performed. This result proves the noisy robustness of the DTC-QRC algorithm at the experimental aspect. 

Near the NTC phase transition boundary ($g = 0.22$), the experimental results consistently underperform compared to those of the DTC-QRC, with a maximum accuracy of up to $81\%$. The maximum accuracy difference between DTC and NTC is $10\%$ in 12 qubits. The enhanced experimental accuracy achieved by DTC relative to NTC may stem from the $\pi$ pairing of eigenstates, conferring greater robustness against coherent and dephasing noise as we discussed the absolute stability of the DTC phase in subsection \ref{stability}.

\section{Discussion and Outlook}

We demonstrate a quantum reservoir computing algorithm employing a many-body localized discrete time crystal as a quantum reservoir, and the algorithm is gradient-free, noise robust,  and scalable. Our results establish a direct connection between quantum phases of matter, non-equilibrium phase transitions, and classification accuracy in quantum machine learning, offering new pathways for harnessing quantum many-body systems in QML and designing hardware-adapted quantum algorithms. The efficiency of the algorithm is demonstrated by its performance, which is comparable to classical multi-layer perceptron under fair comparison. We have analyzed the effect of noise and found that noise has less impact on our algorithm. Experiments in superconducting quantum processors have further demonstrated the noise robustness and efficiency of the algorithm.  The potential classical intractability of the QRC transformation is quantified through the observation of comparative quantum kernel advantage. 

The many-body dynamics in the QRC is achieved using digital quantum simulation, which makes it widely applicable to various NISQ devices. It would also be interesting to investigate the alternative Floquet DTC model for archieving QRC, for example, the quantum many-body scars DTC \cite{ScarsDTC}, Stark DTC \cite{StarkDTC} and etc.  
The physical insights we provide on the correlation between dynamical phase transition, classification performance, and noise robustness are also suitable for different quantum devices. Large-scale experiments on finely tuned quantum devices are expected.
Due to its efficiency, QRC is expected to play an important role in classical and quantum machine learning algorithms to achieve diverse tasks, and has the potential to demonstrate practical quantum enhancement in specific tasks. Beyond that, it's natural to employ QRC to process quantum data, such as the identification of entangled states and classification of the quantum phase.

\begin{acknowledgments}
The authors thank Zhaohui Wei for helpful discussions. 
This work is supported by the Beijing Institute of Technology Research Fund Program under Grant No. 2024CX01015. 
\end{acknowledgments}

\newpage
\appendix
\section{Phase Identification and Sample Efficiency Analysis}
\subsection{Identify the Dynamical Phase Transition }

The key feature of the DTC phase is the long-range spatial order characterized by the Edwards-Anderson spin-glass order parameter \cite{spinglass1975,mblglass2014,spinglass2016}:
\begin{equation}
    \chi^{SG}=\frac{1}{L}\sum_{i\ne j }\left \langle  Z_{i}Z_{j} \right \rangle ^{2}.
\end{equation}
$\chi^{SG}$ vanishes with increasing system size $L$ in the thermal phase, while growing in the NTC and DTC phases. Therefore, the intersection of $\chi^{SG}$ of different $L$ is considered to be the phase transition boundary.

\subsection{Sample Size Analysis \label{sample}}
\begin{figure}[t]
    \centering
    \includegraphics[width=0.95\linewidth]{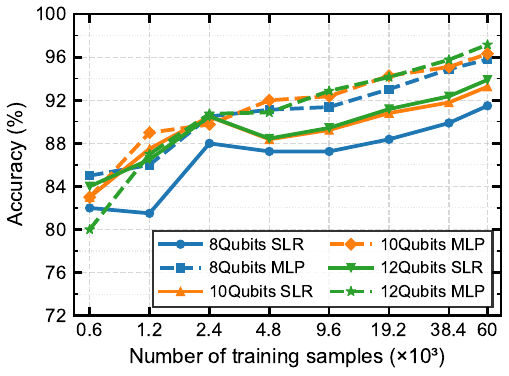}
    \caption{Classification accuracy versus training sample size for angle coding and dense angle coding methods. Performance improves substantially with larger datasets for both simple linear regression and multilayer perceptron. }
    \label{fig:sample}
\end{figure}
To reduce computational costs and match experimental configurations, our main simulations use 1200 training samples and 200 test samples. Here, we investigate how the training sample size affects performance for both angle coding and dense angle coding methods. We implement a MLP for classical post-processing to demonstrate enhanced performance capabilities. The architecture of the MLP is MLP2 in \ref{tab:nn}. The results indicate a significant increase in classification accuracy with larger training sets. As shown in Fig.~\ref{fig:sample} , for SLR, the accuracy exceeds $90 \%$, reaching $93.9\% $ at 12 qubits. For MLP, the maximum accuracy at 12 qubits reaches $97.17\%$. 
\begin{table}[]
    \centering
    \begin{tabular}{l|c|c}
        \hline
        Network & Architecture & Details \\
        \hline
        SLR & Output only (2/10) & Linear classifier \\
        MLP1 & 64-32-16-10 & ReLU + Dropout (0.2) \\
        MLP2 & 128-64-32-10 & ReLU + Dropout (0.3, 0.2, 0.2) \\
        \hline
    \end{tabular}
    \caption{Fully connected neural networks for classification. SLR is the classical post-processing used in the main text; MLP1 is the classical network used for comparison with quantum reservoir; MLP2 is a post-processing network used to show the algorithm's ability to perform better classification in Appendix \ref{sample}.}
    \label{tab:nn}
\end{table}

\section{Noise Model and Robustness analysis}
\begin{figure}[htb]
    \centering
 \includegraphics[width=0.95\linewidth]{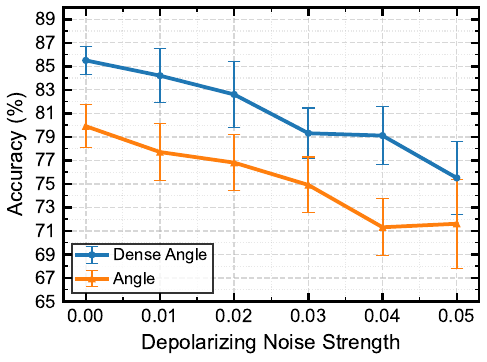}
    \caption{Test accuracy as a function of two-qubit depolarizing noise strength. }
    \label{fig:noisestrength}
\end{figure}
\subsection{Noise Model \label{noisemodel}}
\subsubsection{State Preparation Noise}
 The noise of state preparation is modeled as a probabilistic error in which the intended $\ket{0}$ state initialization has a probability $p_1$ of erroneously preparing the $\ket{1}$ state. This noise mechanism is implemented by applying a Pauli-X gate with $p_1$ immediately after the initialization procedure.
 
\subsubsection{Thermal Relaxation Noise}
Thermal relaxation noise arises from the interaction between qubits and their thermal environment, leading to energy dissipation (characterized by $T_{1}$) and dephasing (governed by $T_{2}$). 
When $T_{2}<T_{1}$ the thermal relaxation channel can be expressed as a probabilistic mixture of reset operations and unitary errors $\left \{ I,Z,Reset \right \} $ with probabilities 

\begin{equation}
\begin{cases}
\begin{aligned}
p_{\text{reset}} &= 1 - e^{-t/T_1} \\
p_{z} &= \frac{1}{2} (1 - p_{\text{reset}}) \left[1 - e^{-t \left( T_2^{-1} - T_1^{-1} \right)} \right] \\
p_{\text{id}} &= 1 - p_z - p_{\text{reset}}
\end{aligned}
\end{cases}
\end{equation}

In the configuration of the noise simulation parameters, to accurately align with the noise characteristics of real quantum devices,the relaxation time $T_{1} \sim \mathcal{N}(\mu = 73\ \mu s,\ \sigma = 10\ \mu s)$, the dephasing time $T_{2} \sim \mathcal{N}(\mu = 25\ \mu s, \ \sigma = 5\ \mu s) $.   Single-qubit gates were configured with $50$ ns operation durations. The $u_{3}$ gate, which is composed of three single qubit gates with $150$ ns operation durations. Two-qubit gate operations exhibited an extended duration of $500$ ns. The measurement operations were modeled as $1 \ \mu s$ delay processes without errors.

\subsubsection{Depolarizing Noise}
One- and two-qubit depolarizing noise models are shown as follows. 
\begin{align}
    \mathcal{N}_{dp}^{1q} &= (1-p_{1})I + \frac{p_{1}}{3} \sum_{\alpha=1}^{3} \sigma_{\alpha}, \\
    \mathcal{N}_{dp}^{2q} &= (1-p_{2})I + \frac{p_{2}}{15} \sum_{\alpha,\beta=1}^{3} \sigma_{\alpha} \sigma_{\beta}
\end{align}
Conservative order-of-magnitude estimates for depolarizing error rates with current technology are $p_1 \approx 10^{-3}$ and $p_2 \approx 10^{-2}$ \cite{arute2019supermarcy,li2023single}. When simulating the phase-dependent test accuracy.

\subsubsection{Depolarizing Noise Effects \label{noisestrength}}
To further investigate the effects of noise on algorithm performance, we perform simulations using depolarizing noise at varying strengths. The test accuracy was observed across two qubit noise strengths from $1\times 10^{-2}$ to $5\times 10^{-2}$ .  With corresponding single-qubit noise strength one order of magnitude lower. Simulations use 600 training samples and 100 test samples, and the number of Floquet cycles is $m=12$. Results in  Fig~\ref{fig:noisestrength} indicate that at state-of-the-art noise levels ($\sim
0.01$), the test accuracy remains virtually unchanged. Under cloud-platform noise configurations, the success rate decreases by approximately 2\%. The algorithm remains effective even under the extreme noise. This result further demonstrates the model's noise robustness. 

\bibliography{ref.bib}

\begin{thebibliography}{66}%
\makeatletter
\providecommand \@ifxundefined [1]{%
 \@ifx{#1\undefined}
}%
\providecommand \@ifnum [1]{%
 \ifnum #1\expandafter \@firstoftwo
 \else \expandafter \@secondoftwo
 \fi
}%
\providecommand \@ifx [1]{%
 \ifx #1\expandafter \@firstoftwo
 \else \expandafter \@secondoftwo
 \fi
}%
\providecommand \natexlab [1]{#1}%
\providecommand \enquote  [1]{``#1''}%
\providecommand \bibnamefont  [1]{#1}%
\providecommand \bibfnamefont [1]{#1}%
\providecommand \citenamefont [1]{#1}%
\providecommand \href@noop [0]{\@secondoftwo}%
\providecommand \href [0]{\begingroup \@sanitize@url \@href}%
\providecommand \@href[1]{\@@startlink{#1}\@@href}%
\providecommand \@@href[1]{\endgroup#1\@@endlink}%
\providecommand \@sanitize@url [0]{\catcode `\\12\catcode `\$12\catcode `\&12\catcode `\#12\catcode `\^12\catcode `\_12\catcode `\%12\relax}%
\providecommand \@@startlink[1]{}%
\providecommand \@@endlink[0]{}%
\providecommand \url  [0]{\begingroup\@sanitize@url \@url }%
\providecommand \@url [1]{\endgroup\@href {#1}{\urlprefix }}%
\providecommand \urlprefix  [0]{URL }%
\providecommand \Eprint [0]{\href }%
\providecommand \doibase [0]{https://doi.org/}%
\providecommand \selectlanguage [0]{\@gobble}%
\providecommand \bibinfo  [0]{\@secondoftwo}%
\providecommand \bibfield  [0]{\@secondoftwo}%
\providecommand \translation [1]{[#1]}%
\providecommand \BibitemOpen [0]{}%
\providecommand \bibitemStop [0]{}%
\providecommand \bibitemNoStop [0]{.\EOS\space}%
\providecommand \EOS [0]{\spacefactor3000\relax}%
\providecommand \BibitemShut  [1]{\csname bibitem#1\endcsname}%
\let\auto@bib@innerbib\@empty
\bibitem [{\citenamefont {LeCun}\ \emph {et~al.}(2015)\citenamefont {LeCun}, \citenamefont {Bengio},\ and\ \citenamefont {Hinton}}]{lecun2015learn}%
  \BibitemOpen
  \bibfield  {author} {\bibinfo {author} {\bibfnamefont {Y.}~\bibnamefont {LeCun}}, \bibinfo {author} {\bibfnamefont {Y.}~\bibnamefont {Bengio}},\ and\ \bibinfo {author} {\bibfnamefont {G.}~\bibnamefont {Hinton}},\ }\bibfield  {title} {\bibinfo {title} {Deep learning},\ }\href {https://doi.org/10.1038/nature14539} {\bibfield  {journal} {\bibinfo  {journal} {Nature}\ }\textbf {\bibinfo {volume} {521}},\ \bibinfo {pages} {436} (\bibinfo {year} {2015})}\BibitemShut {NoStop}%
\bibitem [{\citenamefont {Wang}\ \emph {et~al.}(2023)\citenamefont {Wang}, \citenamefont {Fu}, \citenamefont {Du} \emph {et~al.}}]{wang2023learn}%
  \BibitemOpen
  \bibfield  {author} {\bibinfo {author} {\bibfnamefont {H.}~\bibnamefont {Wang}}, \bibinfo {author} {\bibfnamefont {T.}~\bibnamefont {Fu}}, \bibinfo {author} {\bibfnamefont {Y.}~\bibnamefont {Du}}, \emph {et~al.},\ }\bibfield  {title} {\bibinfo {title} {Scientific discovery in the age of artificial intelligence},\ }\href {https://doi.org/10.1038/s41586-023-06221-2} {\bibfield  {journal} {\bibinfo  {journal} {Nature}\ }\textbf {\bibinfo {volume} {620}},\ \bibinfo {pages} {47} (\bibinfo {year} {2023})}\BibitemShut {NoStop}%
\bibitem [{\citenamefont {Krizhevsky}\ \emph {et~al.}(2017)\citenamefont {Krizhevsky}, \citenamefont {Sutskever},\ and\ \citenamefont {Hinton}}]{krizhevsky2017alexnet}%
  \BibitemOpen
  \bibfield  {author} {\bibinfo {author} {\bibfnamefont {A.}~\bibnamefont {Krizhevsky}}, \bibinfo {author} {\bibfnamefont {I.}~\bibnamefont {Sutskever}},\ and\ \bibinfo {author} {\bibfnamefont {G.~E.}\ \bibnamefont {Hinton}},\ }\bibfield  {title} {\bibinfo {title} {Imagenet classification with deep convolutional neural networks},\ }\href {https://doi.org/10.1145/3065386} {\bibfield  {journal} {\bibinfo  {journal} {Commun. ACM}\ }\textbf {\bibinfo {volume} {60}},\ \bibinfo {pages} {84–90} (\bibinfo {year} {2017})}\BibitemShut {NoStop}%
\bibitem [{\citenamefont {He}\ \emph {et~al.}(2016)\citenamefont {He}, \citenamefont {Zhang}, \citenamefont {Ren},\ and\ \citenamefont {Sun}}]{he2016resnet}%
  \BibitemOpen
  \bibfield  {author} {\bibinfo {author} {\bibfnamefont {K.}~\bibnamefont {He}}, \bibinfo {author} {\bibfnamefont {X.}~\bibnamefont {Zhang}}, \bibinfo {author} {\bibfnamefont {S.}~\bibnamefont {Ren}},\ and\ \bibinfo {author} {\bibfnamefont {J.}~\bibnamefont {Sun}},\ }\bibfield  {title} {\bibinfo {title} {Deep residual learning for image recognition},\ }in\ \href {https://doi.org/10.1109/CVPR.2016.90} {\emph {\bibinfo {booktitle} {2016 IEEE Conference on Computer Vision and Pattern Recognition (CVPR)}}}\ (\bibinfo {year} {2016})\ pp.\ \bibinfo {pages} {770--778}\BibitemShut {NoStop}%
\bibitem [{\citenamefont {Senior}\ \emph {et~al.}(2020)\citenamefont {Senior}, \citenamefont {Evans}, \citenamefont {Jumper} \emph {et~al.}}]{senior2020protein}%
  \BibitemOpen
  \bibfield  {author} {\bibinfo {author} {\bibfnamefont {A.~W.}\ \bibnamefont {Senior}}, \bibinfo {author} {\bibfnamefont {R.}~\bibnamefont {Evans}}, \bibinfo {author} {\bibfnamefont {J.}~\bibnamefont {Jumper}}, \emph {et~al.},\ }\bibfield  {title} {\bibinfo {title} {Improved protein structure prediction using potentials from deep learning},\ }\href {https://doi.org/10.1038/s41586-019-1923-7} {\bibfield  {journal} {\bibinfo  {journal} {Nature}\ }\textbf {\bibinfo {volume} {577}},\ \bibinfo {pages} {706} (\bibinfo {year} {2020})}\BibitemShut {NoStop}%
\bibitem [{\citenamefont {Vaswani}\ \emph {et~al.}(2017)\citenamefont {Vaswani}, \citenamefont {Shazeer}, \citenamefont {Parmar}, \citenamefont {Uszkoreit}, \citenamefont {Jones}, \citenamefont {Gomez}, \citenamefont {Kaiser},\ and\ \citenamefont {Polosukhin}}]{vaswani2017attention}%
  \BibitemOpen
  \bibfield  {author} {\bibinfo {author} {\bibfnamefont {A.}~\bibnamefont {Vaswani}}, \bibinfo {author} {\bibfnamefont {N.}~\bibnamefont {Shazeer}}, \bibinfo {author} {\bibfnamefont {N.}~\bibnamefont {Parmar}}, \bibinfo {author} {\bibfnamefont {J.}~\bibnamefont {Uszkoreit}}, \bibinfo {author} {\bibfnamefont {L.}~\bibnamefont {Jones}}, \bibinfo {author} {\bibfnamefont {A.~N.}\ \bibnamefont {Gomez}}, \bibinfo {author} {\bibfnamefont {L.~u.}\ \bibnamefont {Kaiser}},\ and\ \bibinfo {author} {\bibfnamefont {I.}~\bibnamefont {Polosukhin}},\ }\bibfield  {title} {\bibinfo {title} {Attention is all you need},\ }in\ \href {https://proceedings.neurips.cc/paper_files/paper/2017/file/3f5ee243547dee91fbd053c1c4a845aa-Paper.pdf} {\emph {\bibinfo {booktitle} {Advances in Neural Information Processing Systems}}},\ Vol.~\bibinfo {volume} {30},\ \bibinfo {editor} {edited by\ \bibinfo {editor} {\bibfnamefont {I.}~\bibnamefont {Guyon}}, \bibinfo {editor} {\bibfnamefont {U.~V.}\ \bibnamefont {Luxburg}}, \bibinfo {editor}
  {\bibfnamefont {S.}~\bibnamefont {Bengio}}, \bibinfo {editor} {\bibfnamefont {H.}~\bibnamefont {Wallach}}, \bibinfo {editor} {\bibfnamefont {R.}~\bibnamefont {Fergus}}, \bibinfo {editor} {\bibfnamefont {S.}~\bibnamefont {Vishwanathan}},\ and\ \bibinfo {editor} {\bibfnamefont {R.}~\bibnamefont {Garnett}}}\ (\bibinfo  {publisher} {Curran Associates, Inc.},\ \bibinfo {year} {2017})\BibitemShut {NoStop}%
\bibitem [{\citenamefont {Preskill}(2018)}]{preskill2018nisq}%
  \BibitemOpen
  \bibfield  {author} {\bibinfo {author} {\bibfnamefont {J.}~\bibnamefont {Preskill}},\ }\bibfield  {title} {\bibinfo {title} {Quantum {C}omputing in the {NISQ} era and beyond},\ }\href {https://doi.org/10.22331/q-2018-08-06-79} {\bibfield  {journal} {\bibinfo  {journal} {{Quantum}}\ }\textbf {\bibinfo {volume} {2}},\ \bibinfo {pages} {79} (\bibinfo {year} {2018})}\BibitemShut {NoStop}%
\bibitem [{\citenamefont {{Google Quantum AI and Collaborators}}(2025)}]{acharya2024google}%
  \BibitemOpen
  \bibfield  {author} {\bibinfo {author} {\bibnamefont {{Google Quantum AI and Collaborators}}},\ }\bibfield  {title} {\bibinfo {title} {Quantum error correction below the surface code threshold},\ }\href {https://doi.org/10.1038/s41586-024-08449-y} {\bibfield  {journal} {\bibinfo  {journal} {Nature}\ }\textbf {\bibinfo {volume} {638}},\ \bibinfo {pages} {920} (\bibinfo {year} {2025})}\BibitemShut {NoStop}%
\bibitem [{\citenamefont {Kim}\ \emph {et~al.}(2023)\citenamefont {Kim}, \citenamefont {Eddins}, \citenamefont {Anand} \emph {et~al.}}]{kim2023evidence}%
  \BibitemOpen
  \bibfield  {author} {\bibinfo {author} {\bibfnamefont {Y.}~\bibnamefont {Kim}}, \bibinfo {author} {\bibfnamefont {A.}~\bibnamefont {Eddins}}, \bibinfo {author} {\bibfnamefont {S.}~\bibnamefont {Anand}}, \emph {et~al.},\ }\bibfield  {title} {\bibinfo {title} {Evidence for the utility of quantum computing before fault tolerance},\ }\href {https://doi.org/10.1038/s41586-023-06096-3} {\bibfield  {journal} {\bibinfo  {journal} {Nature}\ }\textbf {\bibinfo {volume} {618}},\ \bibinfo {pages} {500} (\bibinfo {year} {2023})}\BibitemShut {NoStop}%
\bibitem [{\citenamefont {Guo}\ \emph {et~al.}(2024)\citenamefont {Guo}, \citenamefont {Wu}, \citenamefont {Ye} \emph {et~al.}}]{guo2024500}%
  \BibitemOpen
  \bibfield  {author} {\bibinfo {author} {\bibfnamefont {S.~A.}\ \bibnamefont {Guo}}, \bibinfo {author} {\bibfnamefont {Y.~K.}\ \bibnamefont {Wu}}, \bibinfo {author} {\bibfnamefont {J.}~\bibnamefont {Ye}}, \emph {et~al.},\ }\bibfield  {title} {\bibinfo {title} {A site-resolved two-dimensional quantum simulator with hundreds of trapped ions},\ }\href {https://doi.org/10.1038/s41586-024-07459-0} {\bibfield  {journal} {\bibinfo  {journal} {Nature}\ }\textbf {\bibinfo {volume} {630}},\ \bibinfo {pages} {613} (\bibinfo {year} {2024})}\BibitemShut {NoStop}%
\bibitem [{\citenamefont {Bluvstein}\ \emph {et~al.}(2024)\citenamefont {Bluvstein}, \citenamefont {Evered}, \citenamefont {Geim} \emph {et~al.}}]{bluvstein2024logical}%
  \BibitemOpen
  \bibfield  {author} {\bibinfo {author} {\bibfnamefont {D.}~\bibnamefont {Bluvstein}}, \bibinfo {author} {\bibfnamefont {S.~J.}\ \bibnamefont {Evered}}, \bibinfo {author} {\bibfnamefont {A.~A.}\ \bibnamefont {Geim}}, \emph {et~al.},\ }\bibfield  {title} {\bibinfo {title} {Logical quantum processor based on reconfigurable atom arrays},\ }\href {https://doi.org/10.1038/s41586-023-06927-3} {\bibfield  {journal} {\bibinfo  {journal} {Nature}\ }\textbf {\bibinfo {volume} {626}},\ \bibinfo {pages} {58} (\bibinfo {year} {2024})}\BibitemShut {NoStop}%
\bibitem [{\citenamefont {Kandala}\ \emph {et~al.}(2017)\citenamefont {Kandala}, \citenamefont {Mezzacapo}, \citenamefont {Temme} \emph {et~al.}}]{kandala2017vqe}%
  \BibitemOpen
  \bibfield  {author} {\bibinfo {author} {\bibfnamefont {A.}~\bibnamefont {Kandala}}, \bibinfo {author} {\bibfnamefont {A.}~\bibnamefont {Mezzacapo}}, \bibinfo {author} {\bibfnamefont {K.}~\bibnamefont {Temme}}, \emph {et~al.},\ }\bibfield  {title} {\bibinfo {title} {Hardware-efficient variational quantum eigensolver for small molecules and quantum magnets},\ }\href {https://doi.org/10.1038/nature23879} {\bibfield  {journal} {\bibinfo  {journal} {Nature}\ }\textbf {\bibinfo {volume} {549}},\ \bibinfo {pages} {242} (\bibinfo {year} {2017})}\BibitemShut {NoStop}%
\bibitem [{\citenamefont {Cong}\ \emph {et~al.}(2019)\citenamefont {Cong}, \citenamefont {Choi},\ and\ \citenamefont {Lukin}}]{cong2019quantum}%
  \BibitemOpen
  \bibfield  {author} {\bibinfo {author} {\bibfnamefont {I.}~\bibnamefont {Cong}}, \bibinfo {author} {\bibfnamefont {S.}~\bibnamefont {Choi}},\ and\ \bibinfo {author} {\bibfnamefont {M.~D.}\ \bibnamefont {Lukin}},\ }\bibfield  {title} {\bibinfo {title} {Quantum convolutional neural networks},\ }\href {https://doi.org/10.1038/s41567-019-0648-8} {\bibfield  {journal} {\bibinfo  {journal} {Nature Physics}\ }\textbf {\bibinfo {volume} {15}},\ \bibinfo {pages} {1273} (\bibinfo {year} {2019})}\BibitemShut {NoStop}%
\bibitem [{\citenamefont {Herrmann}\ \emph {et~al.}(2022)\citenamefont {Herrmann}, \citenamefont {Llima}, \citenamefont {Remm} \emph {et~al.}}]{herrmann2022qcnn}%
  \BibitemOpen
  \bibfield  {author} {\bibinfo {author} {\bibfnamefont {J.}~\bibnamefont {Herrmann}}, \bibinfo {author} {\bibfnamefont {S.~M.}\ \bibnamefont {Llima}}, \bibinfo {author} {\bibfnamefont {A.}~\bibnamefont {Remm}}, \emph {et~al.},\ }\bibfield  {title} {\bibinfo {title} {Realizing quantum convolutional neural networks on a superconducting quantum processor to recognize quantum phases},\ }\href {https://doi.org/10.1038/s41467-022-31679-5} {\bibfield  {journal} {\bibinfo  {journal} {Nat. Commun.}\ }\textbf {\bibinfo {volume} {13}},\ \bibinfo {pages} {4144} (\bibinfo {year} {2022})}\BibitemShut {NoStop}%
\bibitem [{\citenamefont {Hur}\ \emph {et~al.}(2022)\citenamefont {Hur}, \citenamefont {Kim},\ and\ \citenamefont {Park}}]{hur2022qcnnc}%
  \BibitemOpen
  \bibfield  {author} {\bibinfo {author} {\bibfnamefont {T.}~\bibnamefont {Hur}}, \bibinfo {author} {\bibfnamefont {L.}~\bibnamefont {Kim}},\ and\ \bibinfo {author} {\bibfnamefont {D.~K.}\ \bibnamefont {Park}},\ }\bibfield  {title} {\bibinfo {title} {Quantum convolutional neural network for classical data classification},\ }\href {https://doi.org/10.1007/s42484-021-00061-x} {\bibfield  {journal} {\bibinfo  {journal} {Quantum Mach. Intell.}\ }\textbf {\bibinfo {volume} {4}},\ \bibinfo {pages} {3} (\bibinfo {year} {2022})}\BibitemShut {NoStop}%
\bibitem [{\citenamefont {Farhi}\ \emph {et~al.}(2014)\citenamefont {Farhi}, \citenamefont {Goldstone},\ and\ \citenamefont {Gutmann}}]{farhi2014qaoa}%
  \BibitemOpen
  \bibfield  {author} {\bibinfo {author} {\bibfnamefont {E.}~\bibnamefont {Farhi}}, \bibinfo {author} {\bibfnamefont {J.}~\bibnamefont {Goldstone}},\ and\ \bibinfo {author} {\bibfnamefont {S.}~\bibnamefont {Gutmann}},\ }\bibfield  {title} {\bibinfo {title} {A quantum approximate optimization algorithm},\ }\href {https://arxiv.org/abs/1411.4028} {\bibfield  {journal} {\bibinfo  {journal} {arXiv preprint}\ } (\bibinfo {year} {2014})},\ \Eprint {https://arxiv.org/abs/1411.4028} {arXiv:1411.4028 [quant-ph]} \BibitemShut {NoStop}%
\bibitem [{pan(2023)}]{pan2023deepgradient}%
  \BibitemOpen
  \bibfield  {title} {\bibinfo {title} {Deep quantum neural networks on a superconducting processor},\ }\href {https://doi.org/10.1038/s41467-023-39785-8} {\bibfield  {journal} {\bibinfo  {journal} {Nat. Commun.}\ }\textbf {\bibinfo {volume} {14}},\ \bibinfo {pages} {4006} (\bibinfo {year} {2023})}\BibitemShut {NoStop}%
\bibitem [{\citenamefont {McClean}\ \emph {et~al.}(2018)\citenamefont {McClean}, \citenamefont {Boixo}, \citenamefont {Smelyanskiy} \emph {et~al.}}]{mcclean2018bpqnn}%
  \BibitemOpen
  \bibfield  {author} {\bibinfo {author} {\bibfnamefont {J.~R.}\ \bibnamefont {McClean}}, \bibinfo {author} {\bibfnamefont {S.}~\bibnamefont {Boixo}}, \bibinfo {author} {\bibfnamefont {V.~N.}\ \bibnamefont {Smelyanskiy}}, \emph {et~al.},\ }\bibfield  {title} {\bibinfo {title} {Barren plateaus in quantum neural network training landscapes},\ }\href {https://doi.org/10.1038/s41467-018-07090-4} {\bibfield  {journal} {\bibinfo  {journal} {Nat. Commun.}\ }\textbf {\bibinfo {volume} {9}},\ \bibinfo {pages} {4812} (\bibinfo {year} {2018})}\BibitemShut {NoStop}%
\bibitem [{\citenamefont {Anschuetz}\ and\ \citenamefont {Kiani}(2022)}]{anschuetz2022entanbps}%
  \BibitemOpen
  \bibfield  {author} {\bibinfo {author} {\bibfnamefont {E.~R.}\ \bibnamefont {Anschuetz}}\ and\ \bibinfo {author} {\bibfnamefont {B.~T.}\ \bibnamefont {Kiani}},\ }\bibfield  {title} {\bibinfo {title} {Quantum variational algorithms are swamped with traps},\ }\href {https://doi.org/10.1038/s41467-022-35364-5} {\bibfield  {journal} {\bibinfo  {journal} {Nat. Commun.}\ }\textbf {\bibinfo {volume} {13}},\ \bibinfo {pages} {7760} (\bibinfo {year} {2022})}\BibitemShut {NoStop}%
\bibitem [{\citenamefont {Bittel}\ and\ \citenamefont {Kliesch}(2021)}]{bittel2021vqanp}%
  \BibitemOpen
  \bibfield  {author} {\bibinfo {author} {\bibfnamefont {L.}~\bibnamefont {Bittel}}\ and\ \bibinfo {author} {\bibfnamefont {M.}~\bibnamefont {Kliesch}},\ }\bibfield  {title} {\bibinfo {title} {Training variational quantum algorithms is np-hard},\ }\href {https://doi.org/10.1103/PhysRevLett.127.120502} {\bibfield  {journal} {\bibinfo  {journal} {Phys. Rev. Lett.}\ }\textbf {\bibinfo {volume} {127}},\ \bibinfo {pages} {120502} (\bibinfo {year} {2021})}\BibitemShut {NoStop}%
\bibitem [{\citenamefont {Fujii}\ and\ \citenamefont {Nakajima}(2017)}]{qrc17kohei}%
  \BibitemOpen
  \bibfield  {author} {\bibinfo {author} {\bibfnamefont {K.}~\bibnamefont {Fujii}}\ and\ \bibinfo {author} {\bibfnamefont {K.}~\bibnamefont {Nakajima}},\ }\bibfield  {title} {\bibinfo {title} {Harnessing disordered-ensemble quantum dynamics for machine learning},\ }\href {https://doi.org/10.1103/PhysRevApplied.8.024030} {\bibfield  {journal} {\bibinfo  {journal} {Phys. Rev. Appl.}\ }\textbf {\bibinfo {volume} {8}},\ \bibinfo {pages} {024030} (\bibinfo {year} {2017})}\BibitemShut {NoStop}%
\bibitem [{\citenamefont {Nakajima}\ \emph {et~al.}(2019)\citenamefont {Nakajima}, \citenamefont {Fujii}, \citenamefont {Negoro}, \citenamefont {Mitarai},\ and\ \citenamefont {Kitagawa}}]{qrc19kohei}%
  \BibitemOpen
  \bibfield  {author} {\bibinfo {author} {\bibfnamefont {K.}~\bibnamefont {Nakajima}}, \bibinfo {author} {\bibfnamefont {K.}~\bibnamefont {Fujii}}, \bibinfo {author} {\bibfnamefont {M.}~\bibnamefont {Negoro}}, \bibinfo {author} {\bibfnamefont {K.}~\bibnamefont {Mitarai}},\ and\ \bibinfo {author} {\bibfnamefont {M.}~\bibnamefont {Kitagawa}},\ }\bibfield  {title} {\bibinfo {title} {Boosting computational power through spatial multiplexing in quantum reservoir computing},\ }\href {https://doi.org/10.1103/PhysRevApplied.11.034021} {\bibfield  {journal} {\bibinfo  {journal} {Phys. Rev. Appl.}\ }\textbf {\bibinfo {volume} {11}},\ \bibinfo {pages} {034021} (\bibinfo {year} {2019})}\BibitemShut {NoStop}%
\bibitem [{\citenamefont {Ghosh}\ \emph {et~al.}(2019)\citenamefont {Ghosh}, \citenamefont {Opala}, \citenamefont {Matuszewski} \emph {et~al.}}]{ghosh2019qrcp}%
  \BibitemOpen
  \bibfield  {author} {\bibinfo {author} {\bibfnamefont {S.}~\bibnamefont {Ghosh}}, \bibinfo {author} {\bibfnamefont {A.}~\bibnamefont {Opala}}, \bibinfo {author} {\bibfnamefont {M.}~\bibnamefont {Matuszewski}}, \emph {et~al.},\ }\bibfield  {title} {\bibinfo {title} {Quantum reservoir processing},\ }\href {https://doi.org/10.1038/s41534-019-0149-8} {\bibfield  {journal} {\bibinfo  {journal} {npj Quantum Inf.}\ }\textbf {\bibinfo {volume} {5}},\ \bibinfo {pages} {35} (\bibinfo {year} {2019})}\BibitemShut {NoStop}%
\bibitem [{\citenamefont {Mujal}\ \emph {et~al.}(2023)\citenamefont {Mujal}, \citenamefont {Mart{\'{\i}}nez-Pe{\~n}a}, \citenamefont {Giorgi} \emph {et~al.}}]{mujal2023qrcweak}%
  \BibitemOpen
  \bibfield  {author} {\bibinfo {author} {\bibfnamefont {P.}~\bibnamefont {Mujal}}, \bibinfo {author} {\bibfnamefont {R.}~\bibnamefont {Mart{\'{\i}}nez-Pe{\~n}a}}, \bibinfo {author} {\bibfnamefont {G.~L.}\ \bibnamefont {Giorgi}}, \emph {et~al.},\ }\bibfield  {title} {\bibinfo {title} {Time-series quantum reservoir computing with weak and projective measurements},\ }\href {https://doi.org/10.1038/s41534-023-00682-z} {\bibfield  {journal} {\bibinfo  {journal} {npj Quantum Inf.}\ }\textbf {\bibinfo {volume} {9}},\ \bibinfo {pages} {16} (\bibinfo {year} {2023})}\BibitemShut {NoStop}%
\bibitem [{\citenamefont {Ghosh}\ \emph {et~al.}(2021)\citenamefont {Ghosh}, \citenamefont {Krisnanda}, \citenamefont {Paterek} \emph {et~al.}}]{ghosh2021qrccircuits}%
  \BibitemOpen
  \bibfield  {author} {\bibinfo {author} {\bibfnamefont {S.}~\bibnamefont {Ghosh}}, \bibinfo {author} {\bibfnamefont {T.}~\bibnamefont {Krisnanda}}, \bibinfo {author} {\bibfnamefont {T.}~\bibnamefont {Paterek}}, \emph {et~al.},\ }\bibfield  {title} {\bibinfo {title} {Realising and compressing quantum circuits with quantum reservoir computing},\ }\href {https://doi.org/10.1038/s42005-021-00606-3} {\bibfield  {journal} {\bibinfo  {journal} {Commun. Phys.}\ }\textbf {\bibinfo {volume} {4}},\ \bibinfo {pages} {105} (\bibinfo {year} {2021})}\BibitemShut {NoStop}%
\bibitem [{\citenamefont {Mujal}\ \emph {et~al.}(2021)\citenamefont {Mujal}, \citenamefont {Mart{\'{\i}}nez-Pe{\~n}a}, \citenamefont {Nokkala} \emph {et~al.}}]{mujal2021qrcreview}%
  \BibitemOpen
  \bibfield  {author} {\bibinfo {author} {\bibfnamefont {P.}~\bibnamefont {Mujal}}, \bibinfo {author} {\bibfnamefont {R.}~\bibnamefont {Mart{\'{\i}}nez-Pe{\~n}a}}, \bibinfo {author} {\bibfnamefont {J.}~\bibnamefont {Nokkala}}, \emph {et~al.},\ }\bibfield  {title} {\bibinfo {title} {Opportunities in quantum reservoir computing and extreme learning machines},\ }\href {https://doi.org/10.1002/qute.202100027} {\bibfield  {journal} {\bibinfo  {journal} {Adv. Quantum Technol.}\ }\textbf {\bibinfo {volume} {4}},\ \bibinfo {pages} {2100027} (\bibinfo {year} {2021})}\BibitemShut {NoStop}%
\bibitem [{\citenamefont {Jaeger}(2001)}]{jaeger2001crc}%
  \BibitemOpen
  \bibfield  {author} {\bibinfo {author} {\bibfnamefont {H.}~\bibnamefont {Jaeger}},\ }\bibfield  {title} {\bibinfo {title} {The “echo state” approach to analysing and training recurrent neural networks-with an erratum note},\ }\href@noop {} {\bibfield  {journal} {\bibinfo  {journal} {Bonn, Germany: German national research center for information technology gmd technical report}\ }\textbf {\bibinfo {volume} {148}},\ \bibinfo {pages} {13} (\bibinfo {year} {2001})}\BibitemShut {NoStop}%
\bibitem [{\citenamefont {Maass}\ \emph {et~al.}(2002)\citenamefont {Maass}, \citenamefont {Natschläger},\ and\ \citenamefont {Markram}}]{maass2002crc}%
  \BibitemOpen
  \bibfield  {author} {\bibinfo {author} {\bibfnamefont {W.}~\bibnamefont {Maass}}, \bibinfo {author} {\bibfnamefont {T.}~\bibnamefont {Natschläger}},\ and\ \bibinfo {author} {\bibfnamefont {H.}~\bibnamefont {Markram}},\ }\bibfield  {title} {\bibinfo {title} {Real-time computing without stable states: A new framework for neural computation based on perturbations},\ }\href {https://doi.org/10.1162/089976602760407955} {\bibfield  {journal} {\bibinfo  {journal} {Neural Comput.}\ }\textbf {\bibinfo {volume} {14}},\ \bibinfo {pages} {2531} (\bibinfo {year} {2002})}\BibitemShut {NoStop}%
\bibitem [{\citenamefont {Tanaka}\ \emph {et~al.}(2019)\citenamefont {Tanaka}, \citenamefont {Yamane}, \citenamefont {H{\'e}roux} \emph {et~al.}}]{tanaka2019crc}%
  \BibitemOpen
  \bibfield  {author} {\bibinfo {author} {\bibfnamefont {G.}~\bibnamefont {Tanaka}}, \bibinfo {author} {\bibfnamefont {T.}~\bibnamefont {Yamane}}, \bibinfo {author} {\bibfnamefont {J.~B.}\ \bibnamefont {H{\'e}roux}}, \emph {et~al.},\ }\bibfield  {title} {\bibinfo {title} {Recent advances in physical reservoir computing: A review},\ }\href {https://doi.org/10.1016/j.neunet.2019.03.005} {\bibfield  {journal} {\bibinfo  {journal} {Neural Networks}\ }\textbf {\bibinfo {volume} {115}},\ \bibinfo {pages} {100} (\bibinfo {year} {2019})}\BibitemShut {NoStop}%
\bibitem [{\citenamefont {Yan}\ \emph {et~al.}(2024)\citenamefont {Yan}, \citenamefont {Huang}, \citenamefont {Bienstman} \emph {et~al.}}]{yan2024crc}%
  \BibitemOpen
  \bibfield  {author} {\bibinfo {author} {\bibfnamefont {M.}~\bibnamefont {Yan}}, \bibinfo {author} {\bibfnamefont {C.}~\bibnamefont {Huang}}, \bibinfo {author} {\bibfnamefont {P.}~\bibnamefont {Bienstman}}, \emph {et~al.},\ }\bibfield  {title} {\bibinfo {title} {Emerging opportunities and challenges for the future of reservoir computing},\ }\href {https://doi.org/10.1038/s41467-024-45187-1} {\bibfield  {journal} {\bibinfo  {journal} {Nat. Commun.}\ }\textbf {\bibinfo {volume} {15}},\ \bibinfo {pages} {2056} (\bibinfo {year} {2024})}\BibitemShut {NoStop}%
\bibitem [{\citenamefont {Deutsch}(2018)}]{deutsch2018eth}%
  \BibitemOpen
  \bibfield  {author} {\bibinfo {author} {\bibfnamefont {J.~M.}\ \bibnamefont {Deutsch}},\ }\bibfield  {title} {\bibinfo {title} {Eigenstate thermalization hypothesis},\ }\href {https://doi.org/10.1088/1361-6633/aac9f1} {\bibfield  {journal} {\bibinfo  {journal} {Rep. Prog. Phys}\ }\textbf {\bibinfo {volume} {81}},\ \bibinfo {pages} {082001} (\bibinfo {year} {2018})}\BibitemShut {NoStop}%
\bibitem [{\citenamefont {Mart\'{\i}nez-Pe\~na}\ \emph {et~al.}(2021)\citenamefont {Mart\'{\i}nez-Pe\~na}, \citenamefont {Giorgi}, \citenamefont {Nokkala}, \citenamefont {Soriano},\ and\ \citenamefont {Zambrini}}]{martinez2021qrcdynamical}%
  \BibitemOpen
  \bibfield  {author} {\bibinfo {author} {\bibfnamefont {R.}~\bibnamefont {Mart\'{\i}nez-Pe\~na}}, \bibinfo {author} {\bibfnamefont {G.~L.}\ \bibnamefont {Giorgi}}, \bibinfo {author} {\bibfnamefont {J.}~\bibnamefont {Nokkala}}, \bibinfo {author} {\bibfnamefont {M.~C.}\ \bibnamefont {Soriano}},\ and\ \bibinfo {author} {\bibfnamefont {R.}~\bibnamefont {Zambrini}},\ }\bibfield  {title} {\bibinfo {title} {Dynamical phase transitions in quantum reservoir computing},\ }\href {https://doi.org/10.1103/PhysRevLett.127.100502} {\bibfield  {journal} {\bibinfo  {journal} {Phys. Rev. Lett.}\ }\textbf {\bibinfo {volume} {127}},\ \bibinfo {pages} {100502} (\bibinfo {year} {2021})}\BibitemShut {NoStop}%
\bibitem [{\citenamefont {Xia}\ \emph {et~al.}(2023)\citenamefont {Xia}, \citenamefont {Zou}, \citenamefont {Qiu} \emph {et~al.}}]{xia2023qrcconfigured}%
  \BibitemOpen
  \bibfield  {author} {\bibinfo {author} {\bibfnamefont {W.}~\bibnamefont {Xia}}, \bibinfo {author} {\bibfnamefont {J.}~\bibnamefont {Zou}}, \bibinfo {author} {\bibfnamefont {X.}~\bibnamefont {Qiu}}, \emph {et~al.},\ }\bibfield  {title} {\bibinfo {title} {Configured quantum reservoir computing for multi-task machine learning},\ }\href {https://doi.org/10.1016/j.scib.2023.08.040} {\bibfield  {journal} {\bibinfo  {journal} {Sci. Bull.}\ }\textbf {\bibinfo {volume} {68}},\ \bibinfo {pages} {2321} (\bibinfo {year} {2023})}\BibitemShut {NoStop}%
\bibitem [{\citenamefont {Bravo}\ \emph {et~al.}(2022)\citenamefont {Bravo}, \citenamefont {Najafi}, \citenamefont {Gao},\ and\ \citenamefont {Yelin}}]{bravo2022quantumtimex}%
  \BibitemOpen
  \bibfield  {author} {\bibinfo {author} {\bibfnamefont {R.~A.}\ \bibnamefont {Bravo}}, \bibinfo {author} {\bibfnamefont {K.}~\bibnamefont {Najafi}}, \bibinfo {author} {\bibfnamefont {X.}~\bibnamefont {Gao}},\ and\ \bibinfo {author} {\bibfnamefont {S.~F.}\ \bibnamefont {Yelin}},\ }\bibfield  {title} {\bibinfo {title} {Quantum reservoir computing using arrays of rydberg atoms},\ }\href {https://doi.org/10.1103/PRXQuantum.3.030325} {\bibfield  {journal} {\bibinfo  {journal} {PRX Quantum}\ }\textbf {\bibinfo {volume} {3}},\ \bibinfo {pages} {030325} (\bibinfo {year} {2022})}\BibitemShut {NoStop}%
\bibitem [{\citenamefont {Suprano}\ \emph {et~al.}(2024)\citenamefont {Suprano}, \citenamefont {Zia}, \citenamefont {Innocenti}, \citenamefont {Lorenzo}, \citenamefont {Cimini}, \citenamefont {Giordani}, \citenamefont {Palmisano}, \citenamefont {Polino}, \citenamefont {Spagnolo}, \citenamefont {Sciarrino}, \citenamefont {Palma}, \citenamefont {Ferraro},\ and\ \citenamefont {Paternostro}}]{suprano2024expproper}%
  \BibitemOpen
  \bibfield  {author} {\bibinfo {author} {\bibfnamefont {A.}~\bibnamefont {Suprano}}, \bibinfo {author} {\bibfnamefont {D.}~\bibnamefont {Zia}}, \bibinfo {author} {\bibfnamefont {L.}~\bibnamefont {Innocenti}}, \bibinfo {author} {\bibfnamefont {S.}~\bibnamefont {Lorenzo}}, \bibinfo {author} {\bibfnamefont {V.}~\bibnamefont {Cimini}}, \bibinfo {author} {\bibfnamefont {T.}~\bibnamefont {Giordani}}, \bibinfo {author} {\bibfnamefont {I.}~\bibnamefont {Palmisano}}, \bibinfo {author} {\bibfnamefont {E.}~\bibnamefont {Polino}}, \bibinfo {author} {\bibfnamefont {N.}~\bibnamefont {Spagnolo}}, \bibinfo {author} {\bibfnamefont {F.}~\bibnamefont {Sciarrino}}, \bibinfo {author} {\bibfnamefont {G.~M.}\ \bibnamefont {Palma}}, \bibinfo {author} {\bibfnamefont {A.}~\bibnamefont {Ferraro}},\ and\ \bibinfo {author} {\bibfnamefont {M.}~\bibnamefont {Paternostro}},\ }\bibfield  {title} {\bibinfo {title} {Experimental property reconstruction in a photonic quantum extreme learning machine},\ }\href
  {https://doi.org/10.1103/PhysRevLett.132.160802} {\bibfield  {journal} {\bibinfo  {journal} {Phys. Rev. Lett.}\ }\textbf {\bibinfo {volume} {132}},\ \bibinfo {pages} {160802} (\bibinfo {year} {2024})}\BibitemShut {NoStop}%
\bibitem [{\citenamefont {De~Lorenzis}\ \emph {et~al.}(2025)\citenamefont {De~Lorenzis}, \citenamefont {Casado}, \citenamefont {Estarellas}, \citenamefont {Lo~Gullo}, \citenamefont {Lux}, \citenamefont {Plastina}, \citenamefont {Riera},\ and\ \citenamefont {Settino}}]{de2025qelmimage}%
  \BibitemOpen
  \bibfield  {author} {\bibinfo {author} {\bibfnamefont {A.}~\bibnamefont {De~Lorenzis}}, \bibinfo {author} {\bibfnamefont {M.}~\bibnamefont {Casado}}, \bibinfo {author} {\bibfnamefont {M.}~\bibnamefont {Estarellas}}, \bibinfo {author} {\bibfnamefont {N.}~\bibnamefont {Lo~Gullo}}, \bibinfo {author} {\bibfnamefont {T.}~\bibnamefont {Lux}}, \bibinfo {author} {\bibfnamefont {F.}~\bibnamefont {Plastina}}, \bibinfo {author} {\bibfnamefont {A.}~\bibnamefont {Riera}},\ and\ \bibinfo {author} {\bibfnamefont {J.}~\bibnamefont {Settino}},\ }\bibfield  {title} {\bibinfo {title} {Harnessing quantum extreme learning machines for image classification},\ }\href {https://doi.org/10.1103/PhysRevApplied.23.044024} {\bibfield  {journal} {\bibinfo  {journal} {Phys. Rev. Appl.}\ }\textbf {\bibinfo {volume} {23}},\ \bibinfo {pages} {044024} (\bibinfo {year} {2025})}\BibitemShut {NoStop}%
\bibitem [{\citenamefont {Kornja{\v{c}}a}\ \emph {et~al.}(2024)\citenamefont {Kornja{\v{c}}a}, \citenamefont {Hu}, \citenamefont {Zhao} \emph {et~al.}}]{kornjavca2024analog}%
  \BibitemOpen
  \bibfield  {author} {\bibinfo {author} {\bibfnamefont {M.}~\bibnamefont {Kornja{\v{c}}a}}, \bibinfo {author} {\bibfnamefont {H.~Y.}\ \bibnamefont {Hu}}, \bibinfo {author} {\bibfnamefont {C.}~\bibnamefont {Zhao}}, \emph {et~al.},\ }\bibfield  {title} {\bibinfo {title} {Large-scale quantum reservoir learning with an analog quantum computer},\ }\href {https://arxiv.org/abs/2407.02553} {\bibfield  {journal} {\bibinfo  {journal} {arXiv preprint}\ } (\bibinfo {year} {2024})},\ \Eprint {https://arxiv.org/abs/2407.02553} {arXiv:2407.02553} \BibitemShut {NoStop}%
\bibitem [{\citenamefont {Cimini}\ \emph {et~al.}(2025)\citenamefont {Cimini}, \citenamefont {Sohoni}, \citenamefont {Presutti} \emph {et~al.}}]{cimini2025gbs}%
  \BibitemOpen
  \bibfield  {author} {\bibinfo {author} {\bibfnamefont {V.}~\bibnamefont {Cimini}}, \bibinfo {author} {\bibfnamefont {M.~M.}\ \bibnamefont {Sohoni}}, \bibinfo {author} {\bibfnamefont {F.}~\bibnamefont {Presutti}}, \emph {et~al.},\ }\bibfield  {title} {\bibinfo {title} {Large-scale quantum reservoir computing using a gaussian boson sampler},\ }\href {https://arxiv.org/abs/2505.13695} {\bibfield  {journal} {\bibinfo  {journal} {arXiv preprint}\ } (\bibinfo {year} {2025})},\ \Eprint {https://arxiv.org/abs/2505.13695} {arXiv:2505.13695 [quant-ph]} \BibitemShut {NoStop}%
\bibitem [{\citenamefont {Gong}\ \emph {et~al.}(2025)\citenamefont {Gong}, \citenamefont {Chen}, \citenamefont {Liu} \emph {et~al.}}]{gong2025gbs}%
  \BibitemOpen
  \bibfield  {author} {\bibinfo {author} {\bibfnamefont {S.~Q.}\ \bibnamefont {Gong}}, \bibinfo {author} {\bibfnamefont {M.~C.}\ \bibnamefont {Chen}}, \bibinfo {author} {\bibfnamefont {H.~L.}\ \bibnamefont {Liu}}, \emph {et~al.},\ }\bibfield  {title} {\bibinfo {title} {Enhanced image recognition using gaussian boson sampling},\ }\href {https://arxiv.org/abs/2506.19707} {\bibfield  {journal} {\bibinfo  {journal} {arXiv preprint}\ } (\bibinfo {year} {2025})},\ \Eprint {https://arxiv.org/abs/2506.19707} {arXiv:2506.19707 [quant-ph]} \BibitemShut {NoStop}%
\bibitem [{\citenamefont {Xia}\ \emph {et~al.}(2022)\citenamefont {Xia}, \citenamefont {Zou}, \citenamefont {Qiu} \emph {et~al.}}]{xia2022qrcdy}%
  \BibitemOpen
  \bibfield  {author} {\bibinfo {author} {\bibfnamefont {W.}~\bibnamefont {Xia}}, \bibinfo {author} {\bibfnamefont {J.}~\bibnamefont {Zou}}, \bibinfo {author} {\bibfnamefont {X.}~\bibnamefont {Qiu}}, \emph {et~al.},\ }\bibfield  {title} {\bibinfo {title} {The reservoir learning power across quantum many-body localization transition},\ }\href {https://doi.org/10.1007/s11467-022-1158-1} {\bibfield  {journal} {\bibinfo  {journal} {Front. Phys.}\ }\textbf {\bibinfo {volume} {17}},\ \bibinfo {pages} {33506} (\bibinfo {year} {2022})}\BibitemShut {NoStop}%
\bibitem [{\citenamefont {Yao}\ \emph {et~al.}(2017)\citenamefont {Yao}, \citenamefont {Potter}, \citenamefont {Potirniche},\ and\ \citenamefont {Vishwanath}}]{dtcrigid}%
  \BibitemOpen
  \bibfield  {author} {\bibinfo {author} {\bibfnamefont {N.~Y.}\ \bibnamefont {Yao}}, \bibinfo {author} {\bibfnamefont {A.~C.}\ \bibnamefont {Potter}}, \bibinfo {author} {\bibfnamefont {I.-D.}\ \bibnamefont {Potirniche}},\ and\ \bibinfo {author} {\bibfnamefont {A.}~\bibnamefont {Vishwanath}},\ }\bibfield  {title} {\bibinfo {title} {Discrete time crystals: Rigidity, criticality, and realizations},\ }\href {https://doi.org/10.1103/PhysRevLett.118.030401} {\bibfield  {journal} {\bibinfo  {journal} {Phys. Rev. Lett.}\ }\textbf {\bibinfo {volume} {118}},\ \bibinfo {pages} {030401} (\bibinfo {year} {2017})}\BibitemShut {NoStop}%
\bibitem [{\citenamefont {Zaletel}\ \emph {et~al.}(2023)\citenamefont {Zaletel}, \citenamefont {Lukin}, \citenamefont {Monroe}, \citenamefont {Nayak}, \citenamefont {Wilczek},\ and\ \citenamefont {Yao}}]{dtcrmp}%
  \BibitemOpen
  \bibfield  {author} {\bibinfo {author} {\bibfnamefont {M.~P.}\ \bibnamefont {Zaletel}}, \bibinfo {author} {\bibfnamefont {M.}~\bibnamefont {Lukin}}, \bibinfo {author} {\bibfnamefont {C.}~\bibnamefont {Monroe}}, \bibinfo {author} {\bibfnamefont {C.}~\bibnamefont {Nayak}}, \bibinfo {author} {\bibfnamefont {F.}~\bibnamefont {Wilczek}},\ and\ \bibinfo {author} {\bibfnamefont {N.~Y.}\ \bibnamefont {Yao}},\ }\bibfield  {title} {\bibinfo {title} {Colloquium: Quantum and classical discrete time crystals},\ }\href {https://doi.org/10.1103/RevModPhys.95.031001} {\bibfield  {journal} {\bibinfo  {journal} {Rev. Mod. Phys.}\ }\textbf {\bibinfo {volume} {95}},\ \bibinfo {pages} {031001} (\bibinfo {year} {2023})}\BibitemShut {NoStop}%
\bibitem [{\citenamefont {LaRose}\ and\ \citenamefont {Coyle}(2020)}]{larose2020robustencoding}%
  \BibitemOpen
  \bibfield  {author} {\bibinfo {author} {\bibfnamefont {R.}~\bibnamefont {LaRose}}\ and\ \bibinfo {author} {\bibfnamefont {B.}~\bibnamefont {Coyle}},\ }\bibfield  {title} {\bibinfo {title} {Robust data encodings for quantum classifiers},\ }\href {https://doi.org/10.1103/PhysRevA.102.032420} {\bibfield  {journal} {\bibinfo  {journal} {Phys. Rev. A}\ }\textbf {\bibinfo {volume} {102}},\ \bibinfo {pages} {032420} (\bibinfo {year} {2020})}\BibitemShut {NoStop}%
\bibitem [{\citenamefont {{Khemani}}\ \emph {et~al.}(2019)\citenamefont {{Khemani}}, \citenamefont {{Moessner}},\ and\ \citenamefont {{Sondhi}}}]{Khemani2019}%
  \BibitemOpen
  \bibfield  {author} {\bibinfo {author} {\bibfnamefont {V.}~\bibnamefont {{Khemani}}}, \bibinfo {author} {\bibfnamefont {R.}~\bibnamefont {{Moessner}}},\ and\ \bibinfo {author} {\bibfnamefont {S.~L.}\ \bibnamefont {{Sondhi}}},\ }\bibfield  {title} {\bibinfo {title} {{A Brief History of Time Crystals}},\ }\href@noop {} {\bibfield  {journal} {\bibinfo  {journal} {arXiv e-prints}\ } (\bibinfo {year} {2019})},\ \Eprint {https://arxiv.org/abs/1910.10745} {arXiv:1910.10745 [cond-mat.str-el]} \BibitemShut {NoStop}%
\bibitem [{\citenamefont {Ippoliti}\ \emph {et~al.}(2021)\citenamefont {Ippoliti}, \citenamefont {Kechedzhi}, \citenamefont {Moessner}, \citenamefont {Sondhi},\ and\ \citenamefont {Khemani}}]{ippoliti2021manydtcnoise}%
  \BibitemOpen
  \bibfield  {author} {\bibinfo {author} {\bibfnamefont {M.}~\bibnamefont {Ippoliti}}, \bibinfo {author} {\bibfnamefont {K.}~\bibnamefont {Kechedzhi}}, \bibinfo {author} {\bibfnamefont {R.}~\bibnamefont {Moessner}}, \bibinfo {author} {\bibfnamefont {S.}~\bibnamefont {Sondhi}},\ and\ \bibinfo {author} {\bibfnamefont {V.}~\bibnamefont {Khemani}},\ }\bibfield  {title} {\bibinfo {title} {Many-body physics in the nisq era: Quantum programming a discrete time crystal},\ }\href {https://doi.org/10.1103/PRXQuantum.2.030346} {\bibfield  {journal} {\bibinfo  {journal} {PRX Quantum}\ }\textbf {\bibinfo {volume} {2}},\ \bibinfo {pages} {030346} (\bibinfo {year} {2021})}\BibitemShut {NoStop}%
\bibitem [{\citenamefont {Zhang}\ \emph {et~al.}(2017)\citenamefont {Zhang}, \citenamefont {Hess}, \citenamefont {Kyprianidis} \emph {et~al.}}]{zhang2017trapdtc}%
  \BibitemOpen
  \bibfield  {author} {\bibinfo {author} {\bibfnamefont {J.}~\bibnamefont {Zhang}}, \bibinfo {author} {\bibfnamefont {P.}~\bibnamefont {Hess}}, \bibinfo {author} {\bibfnamefont {A.}~\bibnamefont {Kyprianidis}}, \emph {et~al.},\ }\bibfield  {title} {\bibinfo {title} {Observation of a discrete time crystal},\ }\href {https://doi.org/10.1038/nature21413} {\bibfield  {journal} {\bibinfo  {journal} {Nature}\ }\textbf {\bibinfo {volume} {543}},\ \bibinfo {pages} {217} (\bibinfo {year} {2017})}\BibitemShut {NoStop}%
\bibitem [{\citenamefont {Frey}\ and\ \citenamefont {Rachel}(2022)}]{frey2022superdtc}%
  \BibitemOpen
  \bibfield  {author} {\bibinfo {author} {\bibfnamefont {P.}~\bibnamefont {Frey}}\ and\ \bibinfo {author} {\bibfnamefont {S.}~\bibnamefont {Rachel}},\ }\bibfield  {title} {\bibinfo {title} {Realization of a discrete time crystal on 57 qubits of a quantum computer},\ }\href {https://doi.org/10.1126/sciadv.abm7652} {\bibfield  {journal} {\bibinfo  {journal} {Sci. Adv.}\ }\textbf {\bibinfo {volume} {8}},\ \bibinfo {pages} {eabm7652} (\bibinfo {year} {2022})}\BibitemShut {NoStop}%
\bibitem [{\citenamefont {Randall}\ \emph {et~al.}(2021)\citenamefont {Randall}, \citenamefont {Bradley}, \citenamefont {van~der Gronden} \emph {et~al.}}]{randall2021nvdtc}%
  \BibitemOpen
  \bibfield  {author} {\bibinfo {author} {\bibfnamefont {J.}~\bibnamefont {Randall}}, \bibinfo {author} {\bibfnamefont {C.~E.}\ \bibnamefont {Bradley}}, \bibinfo {author} {\bibfnamefont {F.~V.}\ \bibnamefont {van~der Gronden}}, \emph {et~al.},\ }\bibfield  {title} {\bibinfo {title} {Many-body--localized discrete time crystal with a programmable spin-based quantum simulator},\ }\href {https://doi.org/10.1126/science.abk0603} {\bibfield  {journal} {\bibinfo  {journal} {Science}\ }\textbf {\bibinfo {volume} {374}},\ \bibinfo {pages} {1474} (\bibinfo {year} {2021})}\BibitemShut {NoStop}%
\bibitem [{\citenamefont {{Li}}\ and\ \citenamefont {{Yin}}(2025)}]{DTC-VQE2025}%
  \BibitemOpen
  \bibfield  {author} {\bibinfo {author} {\bibfnamefont {X.}~\bibnamefont {{Li}}}\ and\ \bibinfo {author} {\bibfnamefont {Z.-Q.}\ \bibnamefont {{Yin}}},\ }\bibfield  {title} {\bibinfo {title} {{Improve variational quantum eigensolver by many-body localization}},\ }\href {https://doi.org/10.15302/frontphys.2025.023202} {\bibfield  {journal} {\bibinfo  {journal} {Frontiers of Physics}\ }\textbf {\bibinfo {volume} {20}},\ \bibinfo {pages} {023202} (\bibinfo {year} {2025})}\BibitemShut {NoStop}%
\bibitem [{\citenamefont {Mi}\ \emph {et~al.}(2022{\natexlab{a}})\citenamefont {Mi}, \citenamefont {Ippoliti}, \citenamefont {Quintana} \emph {et~al.}}]{mi2022time}%
  \BibitemOpen
  \bibfield  {author} {\bibinfo {author} {\bibfnamefont {X.}~\bibnamefont {Mi}}, \bibinfo {author} {\bibfnamefont {M.}~\bibnamefont {Ippoliti}}, \bibinfo {author} {\bibfnamefont {C.}~\bibnamefont {Quintana}}, \emph {et~al.},\ }\bibfield  {title} {\bibinfo {title} {Time-crystalline eigenstate order on a quantum processor},\ }\href {https://doi.org/10.1038/s41586-021-04257-w} {\bibfield  {journal} {\bibinfo  {journal} {Nature}\ }\textbf {\bibinfo {volume} {601}},\ \bibinfo {pages} {531} (\bibinfo {year} {2022}{\natexlab{a}})}\BibitemShut {NoStop}%
\bibitem [{\citenamefont {Pal}\ and\ \citenamefont {Huse}(2010)}]{pal2010mbl}%
  \BibitemOpen
  \bibfield  {author} {\bibinfo {author} {\bibfnamefont {A.}~\bibnamefont {Pal}}\ and\ \bibinfo {author} {\bibfnamefont {D.~A.}\ \bibnamefont {Huse}},\ }\bibfield  {title} {\bibinfo {title} {Many-body localization phase transition},\ }\href {https://doi.org/10.1103/PhysRevB.82.174411} {\bibfield  {journal} {\bibinfo  {journal} {Phys. Rev. B}\ }\textbf {\bibinfo {volume} {82}},\ \bibinfo {pages} {174411} (\bibinfo {year} {2010})}\BibitemShut {NoStop}%
\bibitem [{\citenamefont {von Keyserlingk}\ \emph {et~al.}(2016)\citenamefont {von Keyserlingk}, \citenamefont {Khemani},\ and\ \citenamefont {Sondhi}}]{Absolutestability}%
  \BibitemOpen
  \bibfield  {author} {\bibinfo {author} {\bibfnamefont {C.~W.}\ \bibnamefont {von Keyserlingk}}, \bibinfo {author} {\bibfnamefont {V.}~\bibnamefont {Khemani}},\ and\ \bibinfo {author} {\bibfnamefont {S.~L.}\ \bibnamefont {Sondhi}},\ }\bibfield  {title} {\bibinfo {title} {Absolute stability and spatiotemporal long-range order in floquet systems},\ }\href {https://doi.org/10.1103/PhysRevB.94.085112} {\bibfield  {journal} {\bibinfo  {journal} {Phys. Rev. B}\ }\textbf {\bibinfo {volume} {94}},\ \bibinfo {pages} {085112} (\bibinfo {year} {2016})}\BibitemShut {NoStop}%
\bibitem [{\citenamefont {Khemani}\ \emph {et~al.}(2016)\citenamefont {Khemani}, \citenamefont {Lazarides}, \citenamefont {Moessner},\ and\ \citenamefont {Sondhi}}]{spinglass2016}%
  \BibitemOpen
  \bibfield  {author} {\bibinfo {author} {\bibfnamefont {V.}~\bibnamefont {Khemani}}, \bibinfo {author} {\bibfnamefont {A.}~\bibnamefont {Lazarides}}, \bibinfo {author} {\bibfnamefont {R.}~\bibnamefont {Moessner}},\ and\ \bibinfo {author} {\bibfnamefont {S.~L.}\ \bibnamefont {Sondhi}},\ }\bibfield  {title} {\bibinfo {title} {Phase structure of driven quantum systems},\ }\href {https://doi.org/10.1103/PhysRevLett.116.250401} {\bibfield  {journal} {\bibinfo  {journal} {Phys. Rev. Lett.}\ }\textbf {\bibinfo {volume} {116}},\ \bibinfo {pages} {250401} (\bibinfo {year} {2016})}\BibitemShut {NoStop}%
\bibitem [{\citenamefont {Mi}\ \emph {et~al.}(2022{\natexlab{b}})\citenamefont {Mi}, \citenamefont {Sonner}, \citenamefont {Niu} \emph {et~al.}}]{edgerobust}%
  \BibitemOpen
  \bibfield  {author} {\bibinfo {author} {\bibfnamefont {X.}~\bibnamefont {Mi}}, \bibinfo {author} {\bibfnamefont {M.}~\bibnamefont {Sonner}}, \bibinfo {author} {\bibfnamefont {M.~Y.}\ \bibnamefont {Niu}}, \emph {et~al.},\ }\bibfield  {title} {\bibinfo {title} {Noise-resilient edge modes on a chain of superconducting qubits},\ }\href {https://doi.org/10.1126/science.abq5769} {\bibfield  {journal} {\bibinfo  {journal} {Science}\ }\textbf {\bibinfo {volume} {378}},\ \bibinfo {pages} {785} (\bibinfo {year} {2022}{\natexlab{b}})}\BibitemShut {NoStop}%
\bibitem [{\citenamefont {Schmid}\ \emph {et~al.}(2024)\citenamefont {Schmid}, \citenamefont {Penner}, \citenamefont {Yang}, \citenamefont {Glazman},\ and\ \citenamefont {von Oppen}}]{dtcrobust}%
  \BibitemOpen
  \bibfield  {author} {\bibinfo {author} {\bibfnamefont {H.}~\bibnamefont {Schmid}}, \bibinfo {author} {\bibfnamefont {A.-G.}\ \bibnamefont {Penner}}, \bibinfo {author} {\bibfnamefont {K.}~\bibnamefont {Yang}}, \bibinfo {author} {\bibfnamefont {L.}~\bibnamefont {Glazman}},\ and\ \bibinfo {author} {\bibfnamefont {F.}~\bibnamefont {von Oppen}},\ }\bibfield  {title} {\bibinfo {title} {Robust spectral $\ensuremath{\pi}$ pairing in the random-field floquet quantum ising model},\ }\href {https://doi.org/10.1103/PhysRevLett.132.210401} {\bibfield  {journal} {\bibinfo  {journal} {Phys. Rev. Lett.}\ }\textbf {\bibinfo {volume} {132}},\ \bibinfo {pages} {210401} (\bibinfo {year} {2024})}\BibitemShut {NoStop}%
\bibitem [{\citenamefont {Larkin}\ and\ \citenamefont {Ovchinnikov}(1969)}]{larkin1969cotoc}%
  \BibitemOpen
  \bibfield  {author} {\bibinfo {author} {\bibfnamefont {A.~I.}\ \bibnamefont {Larkin}}\ and\ \bibinfo {author} {\bibfnamefont {Y.~N.}\ \bibnamefont {Ovchinnikov}},\ }\bibfield  {title} {\bibinfo {title} {Quasiclassical method in the theory of superconductivity},\ }\href@noop {} {\bibfield  {journal} {\bibinfo  {journal} {Sov Phys JETP}\ }\textbf {\bibinfo {volume} {28}},\ \bibinfo {pages} {1200} (\bibinfo {year} {1969})}\BibitemShut {NoStop}%
\bibitem [{\citenamefont {Li}\ \emph {et~al.}(2017)\citenamefont {Li}, \citenamefont {Fan}, \citenamefont {Wang}, \citenamefont {Ye}, \citenamefont {Zeng}, \citenamefont {Zhai}, \citenamefont {Peng},\ and\ \citenamefont {Du}}]{li2017otoc}%
  \BibitemOpen
  \bibfield  {author} {\bibinfo {author} {\bibfnamefont {J.}~\bibnamefont {Li}}, \bibinfo {author} {\bibfnamefont {R.}~\bibnamefont {Fan}}, \bibinfo {author} {\bibfnamefont {H.}~\bibnamefont {Wang}}, \bibinfo {author} {\bibfnamefont {B.}~\bibnamefont {Ye}}, \bibinfo {author} {\bibfnamefont {B.}~\bibnamefont {Zeng}}, \bibinfo {author} {\bibfnamefont {H.}~\bibnamefont {Zhai}}, \bibinfo {author} {\bibfnamefont {X.}~\bibnamefont {Peng}},\ and\ \bibinfo {author} {\bibfnamefont {J.}~\bibnamefont {Du}},\ }\bibfield  {title} {\bibinfo {title} {Measuring out-of-time-order correlators on a nuclear magnetic resonance quantum simulator},\ }\href {https://doi.org/10.1103/PhysRevX.7.031011} {\bibfield  {journal} {\bibinfo  {journal} {Phys. Rev. X}\ }\textbf {\bibinfo {volume} {7}},\ \bibinfo {pages} {031011} (\bibinfo {year} {2017})}\BibitemShut {NoStop}%
\bibitem [{\citenamefont {Huang}\ \emph {et~al.}(2021)\citenamefont {Huang}, \citenamefont {Broughton}, \citenamefont {Mohseni} \emph {et~al.}}]{huang2021power}%
  \BibitemOpen
  \bibfield  {author} {\bibinfo {author} {\bibfnamefont {H.-Y.}\ \bibnamefont {Huang}}, \bibinfo {author} {\bibfnamefont {M.}~\bibnamefont {Broughton}}, \bibinfo {author} {\bibfnamefont {M.}~\bibnamefont {Mohseni}}, \emph {et~al.},\ }\bibfield  {title} {\bibinfo {title} {Power of data in quantum machine learning},\ }\href {https://doi.org/10.1038/s41467-021-22539-9} {\bibfield  {journal} {\bibinfo  {journal} {Nat. Commun.}\ }\textbf {\bibinfo {volume} {12}},\ \bibinfo {pages} {2631} (\bibinfo {year} {2021})}\BibitemShut {NoStop}%
\bibitem [{\citenamefont {Wang}\ \emph {et~al.}(2021)\citenamefont {Wang}, \citenamefont {Du}, \citenamefont {Luo},\ and\ \citenamefont {Tao}}]{wang2021towards}%
  \BibitemOpen
  \bibfield  {author} {\bibinfo {author} {\bibfnamefont {X.}~\bibnamefont {Wang}}, \bibinfo {author} {\bibfnamefont {Y.}~\bibnamefont {Du}}, \bibinfo {author} {\bibfnamefont {Y.}~\bibnamefont {Luo}},\ and\ \bibinfo {author} {\bibfnamefont {D.}~\bibnamefont {Tao}},\ }\bibfield  {title} {\bibinfo {title} {Towards understanding the power of quantum kernels in the {NISQ} era},\ }\href {https://doi.org/10.22331/q-2021-08-30-531} {\bibfield  {journal} {\bibinfo  {journal} {{Quantum}}\ }\textbf {\bibinfo {volume} {5}},\ \bibinfo {pages} {531} (\bibinfo {year} {2021})}\BibitemShut {NoStop}%
\bibitem [{\citenamefont {Group}(2024)}]{QuafuSQC}%
  \BibitemOpen
  \bibfield  {author} {\bibinfo {author} {\bibfnamefont {Q.~S.~Q.}\ \bibnamefont {Group}},\ }\href {https://quafu-sqc.baqis.ac.cn/home} {\bibinfo {title} {Quafu superconducting quantum computing homepage}} (\bibinfo {year} {2024})\BibitemShut {NoStop}%
\bibitem [{\citenamefont {Maskara}\ \emph {et~al.}(2021)\citenamefont {Maskara}, \citenamefont {Michailidis}, \citenamefont {Ho}, \citenamefont {Bluvstein}, \citenamefont {Choi}, \citenamefont {Lukin},\ and\ \citenamefont {Serbyn}}]{ScarsDTC}%
  \BibitemOpen
  \bibfield  {author} {\bibinfo {author} {\bibfnamefont {N.}~\bibnamefont {Maskara}}, \bibinfo {author} {\bibfnamefont {A.~A.}\ \bibnamefont {Michailidis}}, \bibinfo {author} {\bibfnamefont {W.~W.}\ \bibnamefont {Ho}}, \bibinfo {author} {\bibfnamefont {D.}~\bibnamefont {Bluvstein}}, \bibinfo {author} {\bibfnamefont {S.}~\bibnamefont {Choi}}, \bibinfo {author} {\bibfnamefont {M.~D.}\ \bibnamefont {Lukin}},\ and\ \bibinfo {author} {\bibfnamefont {M.}~\bibnamefont {Serbyn}},\ }\bibfield  {title} {\bibinfo {title} {Discrete time-crystalline order enabled by quantum many-body scars: Entanglement steering via periodic driving},\ }\href {https://doi.org/10.1103/PhysRevLett.127.090602} {\bibfield  {journal} {\bibinfo  {journal} {Phys. Rev. Lett.}\ }\textbf {\bibinfo {volume} {127}},\ \bibinfo {pages} {090602} (\bibinfo {year} {2021})}\BibitemShut {NoStop}%
\bibitem [{\citenamefont {Liu}\ \emph {et~al.}(2023)\citenamefont {Liu}, \citenamefont {Zhang}, \citenamefont {Hsieh}, \citenamefont {Zhang},\ and\ \citenamefont {Yao}}]{StarkDTC}%
  \BibitemOpen
  \bibfield  {author} {\bibinfo {author} {\bibfnamefont {S.}~\bibnamefont {Liu}}, \bibinfo {author} {\bibfnamefont {S.-X.}\ \bibnamefont {Zhang}}, \bibinfo {author} {\bibfnamefont {C.-Y.}\ \bibnamefont {Hsieh}}, \bibinfo {author} {\bibfnamefont {S.}~\bibnamefont {Zhang}},\ and\ \bibinfo {author} {\bibfnamefont {H.}~\bibnamefont {Yao}},\ }\bibfield  {title} {\bibinfo {title} {Discrete time crystal enabled by stark many-body localization},\ }\href {https://doi.org/10.1103/PhysRevLett.130.120403} {\bibfield  {journal} {\bibinfo  {journal} {Phys. Rev. Lett.}\ }\textbf {\bibinfo {volume} {130}},\ \bibinfo {pages} {120403} (\bibinfo {year} {2023})}\BibitemShut {NoStop}%
\bibitem [{\citenamefont {Edwards}\ and\ \citenamefont {Anderson}(1975)}]{spinglass1975}%
  \BibitemOpen
  \bibfield  {author} {\bibinfo {author} {\bibfnamefont {S.~F.}\ \bibnamefont {Edwards}}\ and\ \bibinfo {author} {\bibfnamefont {P.~W.}\ \bibnamefont {Anderson}},\ }\bibfield  {title} {\bibinfo {title} {Theory of spin glasses},\ }\href {https://doi.org/10.1088/0305-4608/5/5/017} {\bibfield  {journal} {\bibinfo  {journal} {J. Phys. F: Met. Phys.}\ }\textbf {\bibinfo {volume} {5}},\ \bibinfo {pages} {965} (\bibinfo {year} {1975})}\BibitemShut {NoStop}%
\bibitem [{\citenamefont {Kj\"all}\ \emph {et~al.}(2014)\citenamefont {Kj\"all}, \citenamefont {Bardarson},\ and\ \citenamefont {Pollmann}}]{mblglass2014}%
  \BibitemOpen
  \bibfield  {author} {\bibinfo {author} {\bibfnamefont {J.~A.}\ \bibnamefont {Kj\"all}}, \bibinfo {author} {\bibfnamefont {J.~H.}\ \bibnamefont {Bardarson}},\ and\ \bibinfo {author} {\bibfnamefont {F.}~\bibnamefont {Pollmann}},\ }\bibfield  {title} {\bibinfo {title} {Many-body localization in a disordered quantum ising chain},\ }\href {https://doi.org/10.1103/PhysRevLett.113.107204} {\bibfield  {journal} {\bibinfo  {journal} {Phys. Rev. Lett.}\ }\textbf {\bibinfo {volume} {113}},\ \bibinfo {pages} {107204} (\bibinfo {year} {2014})}\BibitemShut {NoStop}%
\bibitem [{\citenamefont {Arute}\ \emph {et~al.}(2019)\citenamefont {Arute}, \citenamefont {Arya}, \citenamefont {Babbush} \emph {et~al.}}]{arute2019supermarcy}%
  \BibitemOpen
  \bibfield  {author} {\bibinfo {author} {\bibfnamefont {F.}~\bibnamefont {Arute}}, \bibinfo {author} {\bibfnamefont {K.}~\bibnamefont {Arya}}, \bibinfo {author} {\bibfnamefont {R.}~\bibnamefont {Babbush}}, \emph {et~al.},\ }\bibfield  {title} {\bibinfo {title} {Quantum supremacy using a programmable superconducting processor},\ }\href {https://doi.org/10.1038/s41586-019-1666-5} {\bibfield  {journal} {\bibinfo  {journal} {Nature}\ }\textbf {\bibinfo {volume} {574}},\ \bibinfo {pages} {505} (\bibinfo {year} {2019})}\BibitemShut {NoStop}%
\bibitem [{\citenamefont {Li}\ \emph {et~al.}(2023)\citenamefont {Li}, \citenamefont {Liu}, \citenamefont {Zhao} \emph {et~al.}}]{li2023single}%
  \BibitemOpen
  \bibfield  {author} {\bibinfo {author} {\bibfnamefont {Z.}~\bibnamefont {Li}}, \bibinfo {author} {\bibfnamefont {P.}~\bibnamefont {Liu}}, \bibinfo {author} {\bibfnamefont {P.}~\bibnamefont {Zhao}}, \emph {et~al.},\ }\bibfield  {title} {\bibinfo {title} {Error per single-qubit gate below $10^{-4}$ in a superconducting qubit},\ }\href {https://doi.org/10.1038/s41534-023-00781-x} {\bibfield  {journal} {\bibinfo  {journal} {npj Quantum Inf.}\ }\textbf {\bibinfo {volume} {9}},\ \bibinfo {pages} {111} (\bibinfo {year} {2023})}\BibitemShut {NoStop}%
\end{thebibliography}%

\end{document}